\def\tthdump#1{#1} 
\documentclass[11pt]{article}
\input pdfdef.sty
        \usepackage{color,graphicx}
      \usepackage{lastpage}
      \usepackage{ifthen}
      \usepackage{xspace}
      \usepackage{makeidx}
      \usepackage{amsmath}
      \usepackage[pdflinkmargin=5pt,pdfstartview={FitBH -32768},pdfpagemode=None,plainpages=false]{hyperref}
      \makeindex
      
      \tthdump{\voffset= 0.0 cm}
      \tthdump{\hoffset=-2.5 cm}
      \tthdump{\setlength{\textheight}{21.5 true cm}}
      \tthdump{\setlength{\textwidth}{16.5 true cm}}
      \tthdump{\parindent=30pt}

\def\sz{\scriptsize}

\def\mz{\small} 
\def\nz{\normalsize}
\def\lz{\large}
\def\Lz{\Large}

\def\LLz{\LARGE}
\def\hz{\huge}
\def\Hz{\Huge}

\newif\iftth
\def\tthdump#1{#1}

\iftth 
 
\def\qq{.hspac.}

      \def\yp{y¤¤0039;}
      
     \def\hap{h¤¤0039;}

     \def\nap{n¤¤0039;}
     
     \def\pap{p¤¤0039;}

\def\emm#1{.it.{#1}.eit.}
\def\embm#1{.it.{#1}.eit.}
     \def\min{.btypw.{min}.ett.}
     \def\max{.btypw.{max}.ett.}
     \def\cut{.btypw.{cut}.ett.}

 \def\initial{.btypw.{initial}.ett.}
   \def\final{.btypw.{final}.ett.}

    \def\upo#1{.bupp.{\sz #1}.eupp.}
     \def\up#1{.bupp.{\sz #1}.eupp.}
    \def\suo#1{.bsub.{\sz #1}.esub.}
    \def\sub#1{.bsub.{\sz #1}.esub.}
   
   \def\abs#1{¤¤0124;{#1}¤¤0124;} 
\def\absu#1#2{¤¤0124;{#1}¤¤0124;\upo{#2}}
\def\abss#1#2{¤¤0124;{#1}¤¤0124;\suo{#2}}
\fi
\iftth

      \def\cala{\lz .bf.{A}./bf.}

      \def\cald{\lz .bf.{D}./bf..efon.}

      \def\call{\lz .bf.{L}./bf..efon.}
  \def\lapt#1{\lz .bf.{L}./bf.{\Lz ¤¤0123;}{#1}{\Lz ¤¤0125;}.efon..efon..efon.}
      \def\caln{\lz .bf.{N}./bf..efs.}
      
  \def\calds#1{{\lz .bf.{D}./bf.}\suo{#1}}

  \def\calns#1{{\lz .bf.{N}./bf.}\suo{#1}}

   \def\calcsj{{\lz .bf.{C}./bf.}\suo{j}}
   \def\calwsj{{\lz .bf.{W}./bf.}\suo{j}}
   \def\caldsk{{\lz .bf.{D}./bf.}\suo{k}}
   \def\calnsk{{\lz .bf.{N}./bf.}\suo{k}} 
    
\fi
                  \iftth 

\let\otimx={¤¤8855;}
       \def\otimes{\otimx} 
\def\approx{.hspac.¤¤8776;.hspac.}
\def\app{.hspac.¤¤8776;.hspac.}

\def\propaux{¤¤8733;}

        \let\propto=\prop

      \def\bep{¤¤0946;¤¤0039;}
     \def\gamp{¤¤0947;¤¤0039;}
      
        \def\a{¤¤0945;}
       \def\al{¤¤0945;}
    \def\alpha{¤¤0945;}

      \def\eta{¤¤0951;}

        \def\b{¤¤0946;}
       \def\be{¤¤0946;}
     \def\beta{¤¤0946;}
    \def\bes#1{¤¤0946;\suo{#1}}

        \def\g{¤¤0947;}
       
    \def\gamma{¤¤0947;} 
 \def\\gammaux{¤¤0947;} 

        \def\d{¤¤0948;}
       \def\de{¤¤0948;}
    \def\delta{¤¤0948;}
    \def\des#1{¤¤0948;\suo{#1}}

        \def\e{¤¤1297;}
       \def\ep{¤¤1297;}
  \def\epsilon{¤¤1297;}
 \def\epsilaux{¤¤1297;}
    \def\eps#1{¤¤1297;\suo{#1}}

    \def\epsi{\epsilaux\suo{i}}

  \def\lambaux{¤¤0955;}
      \def\lam{¤¤0955;}
     
   \def\lambda{¤¤0955;}
    
   \def\lams#1{¤¤0955;\suo{#1}}
  \def\lambs#1{¤¤0955;\suo{#1}}

\def\lamsu#1#2{¤¤0955;\supot{#1}{#2}}

  \def\Lambaux{\lz{¤¤0923;}.efon.}

   \def\Lambda{\lz{¤¤0923;}.efon.}

    \def\piaux{¤¤0960;}
        
       \def\pi{¤¤0960;}
       \def\Pi{¤¤0928;}
      
 \def\phi{¤¤0966;}
 \def\Phi{¤¤0934;}

      \def\chi{¤¤0962;}
   \def\chiaux{¤¤0962;}

      \def\lamc{\lz{¤¤0923;}.efon.}

       \def\om{¤¤1120;}
      \def\ome{¤¤1120;}
    \def\oms#1{¤¤1120;\suo{#1}}
    \def\omu#1{¤¤1120;\upo{#1}}

      \def\psi{¤¤1137;}
   \def\psis#1{¤¤1137;\suo{#1}}

        \def\t{¤¤0964;}
      \def\tau{¤¤0964;}
   \def\tauaux{¤¤0964;}

    \def\kappa{¤¤0954;}
       
                  \fi
\iftth
   \def\fe#1{\upo{\sz #1}Fe}

       \def\xi{¤¤0958;}
    \def\xiaux{¤¤0958;}
       \def\Xi{¤¤0926;}
      
    \def\Xiaux{¤¤0926;}

       \def\mu{¤¤0956;}

       \def\nu{¤¤0957;}

        \def\n{¤¤0957;}

      \def\delc{\lz{¤¤0916;}.efon.}
     
    \def\deltac{\lz{¤¤0916;}.efon.}
   \def\delcs#1{\lz{¤¤0916;}.efon.\suo{#1}}

\def\gt{¤¤0062;}

\def\le{¤¤8804;}

      \def\rho{¤¤0961;}

    \def\Omega{\lz{¤¤0937;}.efon.}
             \def\tn{{\bf TNG}}
            \def\tng{{\bf TNG}}

            \def\exm{EXM}
            \def\EXM{EXM}
            \def\oxm{EXM}
            \def\OXM{EXM}
            \def\DMF{DMF}
            \def\DFM{DMF}

             \def\tran{{\bf TRANSNU}}
             \def\trans{{\bf TRANSNU}}

            \def\ripl-ii{RIPL-2}
            \def\ripl{RIPL-2}
            \def\ripl2{RIPL-2}

            \def\ripl-iis{{RIPL-2} }
            
            \def\ripl2s{{RIPL-2} }

            \def\ripl-iib{{RIPL-2} }
            
            \def\ripl2b{{RIPL-2} }

 \def\pe{PE}
 \def\cn{CN}
 \def\qe{QE}

\def\cns{CN }

\fi
\iftth 
\def\vsp#1{}
\def\vsm#1{}

     \def\vst{.hspac..hspac..hspac.}

   \def\hsp#1{}
    \def\hstp{.hspac..hspac..hspac..hspac..hspac..hspac.} 

  \def\hst{.no30.}

 \def\vhst{.no30.}

\else 
\def\hsp#1{\hspace{#1}}

\def\vsp#1{\vspace{#1}}
\def\vsm#1{\vspace{-#1}}

\def\vst{\vspace{10pt}}

  \def\hst{\hspace{-30pt}}

 \def\vhst{\vspace{10pt}\hspace{-30pt}}

\def\hstp{\hspace{30pt}}

\fi  
\iftth                  

   \def\bcs#1{B.bsub.{#1}.esub.}

   \def\fcs#1{F.bsub.{#1}.esub.}

   \def\mcs#1{M.bsub.{#1}.esub.}
   \def\mcu#1{M.bupp.{#1}.eupp.}
   \def\ncs#1{N.bsub.{#1}.esub.}

   \def\ocu#1{O.bupp.{#1}.eupp.}
   \def\pcs#1{P.bsub.{#1}.esub.}

   \def\scs#1{S.bsub.{#1}.esub.}

   \def\ucs#1{U.bsub.{#1}.esub.}
   
   \def\vcs#1{V.bsub.{#1}.esub.}

\def\vcus#1#2{V.bupp.{#1}.eupp..bsub.{#2}.esub.}
\def\vcsu#1#2{V.bsub.{#1}.esub..bupp.{#2}.eupp.}

      \def\mcp{M¤¤0039;}

      \def\ucp{U¤¤0039;}

   \def\fcps#1{F¤¤0039;.bsub.{#1}.esub.}

   \def\fcps#1{F¤¤0039;.bsub.{#1}.esub.}

\fi 
\iftth 

     \def\asl{\as{l}}
    \def\asnu{\as{\nu}}
    
   \def\auspl{a\upo{¤¤2009;}\suo{l}}

  \def\auspnu{\as{\nu}\upo{¤¤2009;}}

    \def\bsmu{\bs{\mu}}

  \def\buspmu{\bs{\mu}\upo{¤¤2009;}}

    \def\xspl{x\suo{pl}}
   \def\xspnu{x\suo{p\nu}}
     
   \def\xshmu{x\suo{h\mu}}

    \def\ucsk{U\suo{k}}
    \def\fcsh{F\suo{h}}
    
    \def\fcsp{F\suo{p}}
   \def\fcpsh{F¤¤0039;\suo{h}}
   
   \def\fcpsp{F¤¤0039;\suo{p}}

\fi
\iftth 
\let\logaux={log}
\def\log{\logaux}
       \def\cdotaux={¤¤8230;} 
\def\cdots{\cdotaux}

     \def\sum={¡¡¡¡¡¡¡¡¡¡¡¡¡¡¡¡¡¡¡¡¡¡¡¡¡¡¡¡¡¡¡¡¡¡¡.bfs.{\LLz ¤¤8721;}£££££££££££££££££££££££££££££££££££££££££££££££££££££££££££££.efs.}
\def\sigsu#1#2{sssssssssssssssssssssssssssssssssss{#2}oooooooooooooooooooo{\LLz ¤¤8721;}++++++++++++++++++++++++++subfontsizemo{{#1}}desubdcccccccccccccccccccc}    

\def\sigsil#1{.bfs.{\Lz ¤¤8721;}.efs.\suo{{#1}}}
\def\sumsu#1#2{sssssssssssssssssssssssssssssssssss{#2}oooooooooooooooooooo{\LLz ¤¤8721;}++++++++++++++++++++++++++subfontsizemo{({#1})}desubdcccccccccccccccccccc}
\def\sums#1{¡¡¡¡¡¡¡¡¡¡¡¡¡¡¡¡¡¡¡¡¡¡¡¡¡¡¡¡¡¡¡¡¡¡¡.bfs.{\LLz ¤¤8721;}£££££££££££££££££££££££££££££££££££££££££££££££££££££££££££££.efs.\suo{({#1})}¢¢¢¢¢¢¢¢¢¢¢¢¢¢¢¢¢¢¢¢}
       
   \def\sumsne#1{.bfs.{\Lz ¤¤8721;}.efs.\suo{({#1})}}

\def\sumst#1#2{sssstttttttttttttttttt.bfs.{\LLz ¤¤8721;}.efs.ooootttttttttttttttt¤¤9581;¢axbra¢{\sz ({#1})}¤¤9582;++++++++++++++++++++++++++subfontsizemo¤¤9584;{\sz ({#2})}~¤¤9583;desubdcccccccccccccccccccc}
\let\sigss\sigst
\let\sumss\sumst

\let\sigssu\sumssu

             \def\intaux={\hz ¤¤8747;}
\def\int{iiiiiiiiiiiiiiiiiiiiiiiiiiiiiiiiiii{}oooooooooooooooooooo{\hz ¤¤8747;}++++++++++++++++++++++++++subfontsizemo{}desubdcccccccccccccccccccc}
\def\intsu#1#2{iiiiiiiiiiiiiiiiiiiiiiiiiiiiiiiiiii{\sz #2}oooooooooooooooooooo{\hz ¤¤8747;}++++++++++++++++++++++++++subfontsizemo{{.hspac..hspac..hspac.{\sz #1}}}desubdcccccccccccccccccccc}
       \def\ointaux={\hz ¤¤8750;}
       
              \def\dointaux={{\Hz ¤¤8751;.sub.}}
       \def\doint{\dointaux.ispac..ispac..ispac.}
\def\infty={¤¤8734;} 
\def\prod={\Lz ¤¤8719;}

\let\in=\belong
\let\bel=\belong

   \def\dif{¤¤8800;}
  
\ifmmode
\def\prodsu#1#2{sssssssssssssssssssssssssssssssssss{#2}oooooooooooooooooooo{\LLz ¤¤8719;}++++++++++++++++++++++++++subfontsizemo{({#1})}desubdcccccccccccccccccccc}
\else
\def\prodsu#1#2{sssssssssssssssssssssssssssssssssss${#2}$oooooooooooooooooooo{\LLz ¤¤8719;}++++++++++++++++++++++++++subfontsizemo{(${#1}$)}desubdcccccccccccccccccccc}
\fi

\def\prods#1{¡¡¡¡¡¡¡¡¡¡¡¡¡¡¡¡¡¡¡¡¡¡¡¡¡¡¡¡¡¡¡¡¡¡¡.bfs.{\LLz ¤¤8719;}£££££££££££££££££££££££££££££££££££££££££££££££££££££££££££££.efs.\suo{({#1})}¢¢¢¢¢¢¢¢¢¢¢¢¢¢¢¢¢¢¢¢}
    \def\as#1{{a\suo{#1}}}
    
 \def\asu#1#2{{a\supot{#1}{#2}}}

 \def\aus#1#2{{a\upsot{#1}{#2}}}
   \def\aas#1{{a\suo{#1}}}

    \def\bs#1{{b\suo{#1}}}

 \def\bus#1#2{{b\upsot{#1}{#2}}}

   \def\cas#1{{c\suo{#1}}}

\def\dasu#1#2{{d\supot{#1}{#2}}}

   \def\das#1{{d\suo{#1}}}
   
    \def\eu#1{{\tt e}\upo{#1}}
   \def\eub#1{{\tt e}\upo{(#1)}}

   \def\fas#1{f\suo{#1}}

\def\hasu#1#2{h\suo{#1}\upo{#2}}

   \def\ias#1{{\tt i}\suo{#1}}

   \def\jas#1{{\tt j}\suo{#1}}

    \def\ks#1{{\tt k}\suo{#1}}

   \def\kas#1{{\tt k}\suo{#1}}

    \def\ls#1{{\tt l}\suo{#1}}

    \def\ms#1{{{\tt m}\suo{#1}}}
  
   \def\mas#1{{\tt m}\suo{#1}}

    \def\ns#1{{n\suo{#1}}}

   \def\nus#1{.bfs.n.efs.\suo{#1}}
    \def\nusi{.bfs.n.efs.\suo{i}}
   \def\nas#1{n\suo{#1}}

    \def\pp#1{p\upo{#1}}

    \def\rs#1{r\suo{#1}}

   \def\ras#1{r\suo{#1}}

    \def\ss#1{{s\suo{#1}}}

   \def\sas#1{{s\suo{#1}}} 
   
    \def\us#1{u\suo{#1}}

    \def\xs#1{x\suo{#1}}
    \def\xu#1{x\upo{#1}}

   \def\xas#1{x\suo{#1}}
   \def\xau#1{x\upo{#1}}

    \def\yu#1{y\upo{#1}}

   \def\yau#1{y\upo{#1}}

\fi
\iftth

      \def\da{d}
      
      \def\fa{f}
      \def\ga{{\tt g}}
      \def\ha{h}
      \def\ia{i}
      \def\ja{{\tt j}}
      \def\ka{k}
      \def\la{l}

      \def\na{n}
      
      \def\pa{p}
      
      \def\ra{r}
      \def\sa{s}
      \def\ta{t}
      \def\ua{u}

      \def\xa{x}
      \def\ya{y}
      
\def\obc{\begingroup\bf }
\def\oec{\endgroup}
       \def\ac{A}
       
       \def\cc{C}

       \def\fc{F}

       \def\mc{M}
       \def\nc{N}
       \def\oc{O}
       \def\pc{P}

       \def\sc{S}
       \def\su{S}
       \def\sl{s}
       
       \def\uc{U}
       \def\vc{V}
       
       \def\xc{X}

\fi
\iftth 

\def\widetilde#1{{\it #1}¤¤0771;}

\def\vec#1{¤¤8407;{\it #1}}

\def\hbar{¤¤8463;}

       \def\rbk#1#2{{(}{{#1}¤¤0124;}{{#2}{)}}}

\let\braket\ddbraket

        \def\dbko#1{.bb..fon.{#1}.ek.}
         
     \def\dbkos#1#2{.bb..fon.{#1}.ek..bsub.{#2}.esub.}
      
    \def\dbok#1#2#3{.bb..fon.{#1}¤¤0124;{#2}¤¤0124;{#3}.ek.}

\def\dbokt#1#2#3#4{{.bb..fon.{#1}¤¤0124;{#2}¤¤0124;{#3}.ek..bb..fon.{#3}¤¤0124;{#4}¤¤0124;{#1}.ek.}}
\def\dirackp{\hspace{4pt}}
\def\diracbp{\fop}
  \def\ket#1{.bk.{#1}.ek.}
 \def\dket#1{.bk.{#1}.ek.}

 \def\dbra#1{.bb.{#1}.eb.}

       \def\dkb#1#2{.bk.{#1}.ek..bb.{#2}.eb.}
\let\ketbra\dkb
\def\timesaux{¤¤9747;}
   \def\times{¤¤9747;}  
\def\dagger{¤¤2009;}
\def\dagaux{¤¤2009;}
   \def\dag{¤¤2009;}  
\def\staraux{¤¤0042;}
   
\def\circaux{¤¤2662;}
   
\fi
\iftth

\fi
\iftth
        \def\of#1{{\lz ¤¤0040;}{#1}{\lz ¤¤0041;}}

      \def\brac#1{{\Lz ¤¤0040;}{#1}{\Lz ¤¤0041;}}

   \def\bracs#1#2{{\Lz ¤¤0040;}{#1}{\Lz ¤¤0041;}_{#2}}

      \def\brak#1{{\Lz ¤¤0091;}{#1}{\Lz ¤¤0093;}}

      \def\brar#1{{\Lz ¤¤0123;}{#1}{\Lz ¤¤0125;}}
   \def\brars#1#2{{\Lz ¤¤0123;}{#1}{\Lz ¤¤0125;}_{#2}}

   \def\bla#1#2{{\lz #1}{#2}}
   \def\bra#1#2{{#2}{\lz #1}}

   \def\xps#1{x¤¤0039;\suo{#1}}
      \def\nn{.hspac..hspac.}

      \def\hp{h¤¤0039;}
      \def\pp{p¤¤0039;}
      \def\xp{x¤¤0039;}
      \def\yp{y¤¤0039;}

     \def\cap{c¤¤0039;}

   \def\xps#1{x¤¤0039;\suo{#1}}

   \def\xpu#1{x¤¤0039;\upo{#1}}

   \def\ypu#1{y¤¤0039;\upo{#1}}

\def\emm#1{{\em #1}}

\def\rightleftarrow{.rlarow.}
\fi

\def\suchaux{\mbox{\lz\hspace{1pt}{\begingroup/\endgroup}\hspace{-2pt}:\hspace{3pt}}} 
    
\def\suchthat{\mathe{\suchaux}}

\newcounter{linkp}
\tthdump{}  

\tthdump{}

\iftth
 
\def\cite#1{.bcit.{#1}.ecit.}

\let\ref=\cite
\else
\let\ref=\cite

\fi
\iftth
\def\ret#1{.bf..oah.{#1}.ncite.{#1}.qgt.\theequationppp.cas../bf.}
\def\reg#1{.bf..oah.{#1}.ncite.{#1}.qgt.\theequationppp.cas../bf.}
\def\ree#1{.bf.(.oah.{#1}.ncite.{#1}.qgt.\theequationppp.cas.)./bf.} 
\else
\def\ret#1{\ref{#1}}
\def\reg#1{\ref{#1}}
\def\ree#1{(\ref{#1})}
\fi

\iftth 
 
\else 
 
\fi 
\def\nline{
\begingroup
\iftth
\vskip 0.5cm
\else
\hskip-40cm.
\newline
\parindent=0pt
\fi
\endgroup}
\def\obcent{\begin{center}}
\def\oecent{\end{center}}
\def\obc{\begingroup\bf }
\def\oec{\endgroup}
\def\title#1{\begingroup   \nline\obcent\LLz{{\bf #1}}\oecent\tthdump{\vskip 0.5cm}\endgroup}
\def\author#1{\begingroup        \obcent\nz{ #1}\oecent\endgroup}
\def\address#1{\begingroup \nline\obcent\nz{{\em #1}}\oecent\vskip 1.0cm\endgroup}
\def\listcard{
\let\prevleng\textheight
\setlength{\textheight}{23.0 true cm}
\newpage
\setcounter{section}{97}
\listofcards
\prevleng}
\iftth
\def\listcardname{
\LARGE{\bf List of Cards}
}
\def\listofcards{\label{List of Cards}\section*{\listcardname}\input{/home/fbraga/pctex/texinput/temp/firstlochtm}}
\fi
\iftth 
\newcounter{appendico}
\newcounter{subsubsection}[subsection]
\newcounter{subsssection}[subsubsection]
   \newcounter{paragraph}[subsubsection]
\renewcommand\thesubsubsection{\thesubsection .\@arabic\c@subsubsection} 
\renewcommand\thesubsssection {\thesubsubsection.\@arabic\c@subsssection}
\renewcommand\theparagraph{Sec.\arabic{section}.\arabic{subsection}.\arabic{subsubsection}.\@arabic\c@subsssection}
\def\secto#1{\setcounter{equationp}{0}\setcounter{equation}{0} 
\section{\arabic{section}..hspac.#1}
.oa.section.\arabic{section}.ca.
.oa.sectionn.\arabic{section}.ca.
} 
\def\subsecto#1{\setcounter{subsubsection}{0}\addtocounter{subsection}{1}
{\Lz{\bf \arabic{section}.\arabic{subsection}.hspac.#1}}
.oa.section.\arabic{section}.\arabic{subsection}.ca.
.oa.sectionn.\arabic{section}.\arabic{subsection}.ca.
.oa.Sec.\arabic{section}.\arabic{subsection}.ca.
}
\def\subsubsecto#1{\setcounter{subsssection}{0}\addtocounter{subsubsection}{1}
{\lz{\bf \arabic{section}.\arabic{subsection}.\arabic{subsubsection}.hspac.#1}}
.oa.section.\arabic{section}.\arabic{subsection}.\arabic{subsubsection}.ca.
.oa.sectionn.\arabic{section}.\arabic{subsection}.\arabic{subsubsection}.ca.
.oa.Sec.\arabic{section}.\arabic{subsection}.\arabic{subsubsection}.ca.
}

\def\subsssecto#1{\addtocounter{subsssection}{1}
{\bf \arabic{section}.\arabic{subsection}.\arabic{subsubsection}.\arabic{subsssection}.hspac.#1}
.oa.section.\arabic{section}.\arabic{subsection}.\arabic{subsubsection}.\arabic{subsssection}.ca.
.oa.sectionn.\arabic{section}.\arabic{subsection}.\arabic{subsubsection}.\arabic{subsssection}.ca.
.oa.Sec.\arabic{section}.\arabic{subsection}.\arabic{subsubsection}.\arabic{subsssection}.ca.
}

\def\paragraph#1{\addtocounter{subsssection}{1}
{\bf \arabic{section}.\arabic{subsection}.\arabic{subsubsection}.\arabic{subsssection}.hspac.#1}
.oa.section.\arabic{section}.\arabic{subsection}.\arabic{subsubsection}.\arabic{subsssection}.ca.
.oa.sectionn.\arabic{section}.\arabic{subsection}.\arabic{subsubsection}.\arabic{subsssection}.ca.
.oa.Sec.\arabic{section}.\arabic{subsection}.\arabic{subsubsection}.\arabic{subsssection}.ca.
} 
\fi

\newcounter{scaleadd}
\iftth
\def\deftab#1{
ttatt.oa.table.\arabic{section}t#1.ca.
.oa.table.\arabic{appendico}a\arabic{section}table#1.ca.
}
\else 
\def\deftab#1{}
\fi 
\iftth          
 
\else 
 
\fi
\iftth 
\def\testnum#1#2#3{ 
\ifnum #1>#2{#3}\fi
}
\else
\def\testnum#1#2{} 
\fi 
\iftth          
 
\else 
 
\fi
\iftth          
\def\makfigmm#1#2#3#4#5#6#7{  
\begin{figure}
.oa.figure.\arabic{section}f#1.ca.
.oa.figure.\arabic{appendico}a\arabic{section}figure#1.ca.
.oa.figure.\Roman{appendico}figure#1.ca.
\label{figure.#1}
\label{figuren.#1}
\vspace{#6}
\centerline{
.oa.figure.\Roman{appendico}f#1.ca.
.oat.{#2}.jpg.tit.File '{#2}': {#2}.jpg.cs.
\includegraphics[scale= .55]{#2}
.cas.
}
\begin{card}   
       \def\thecard{Fig.#1}
\ccapl{}
          \label{figure.#1}
           \def\thecard{#1}
\ccapl{}
         \label{figuren.#1}
\end{card}
\vspace{0.55cm}
.pc.\mbox{\bf Fig.#1 #4}
\end{figure}
} 
\else 
\def\makfigmm#1#2#3#4#5#6#7{
\begingroup
\ifpdf
\vspace{0.41cm}
\fi
\vspace{0.5cm}
\hspace{#5}
\begin{figure}
\centerline{
\HideDisplacementBoxes
\ifpdf
\includegraphics[scale= .#3]{#2}
\else
\setcounter{scaleadd}{#3}
\addtocounter{scaleadd}{200}
\BoxedEPSF{#2.ps scaled \arabic{scaleadd}}
\ForceHeight{10cm}
\fi
\ifpdf
\vspace{0.41cm}
\fi
}
\vspace{#6} 
\capto{#4} 
\end{figure}\nulin{}} 
\fi

\def\brkn#1{\begingroup
\tthdump{\newpage\parindent=0pt \hbox to\textwidth{#1}}
\endgroup}
\def\brkl#1{\begingroup
\tthdump{\hbox to\textwidth{#1}}
\endgroup
}
 
\def\linef#1{\begingroup
\tthdump{\parindent=0pt \hbox to\textwidth{#1}}
\endgroup
\noindent
}
\def\brko#1{\begingroup
\tthdump{\parindent=0pt \hbox to\textwidth{\hspace{30pt}#1}}
\endgroup}

      \def\brkm#1{\begingroup
\tthdump{\hbox to\textwidth{#1}}
\endgroup}
\def\brkk#1#2{\begingroup
\tthdump{\hbox to\textwidth{#1}\parindent=0pt \newpage{#2}}
\endgroup}
\def\brkkfirst#1#2{\begingroup
\tthdump{\hbox to\textwidth{\hspace{30pt}#1}\parindent=0pt \newpage{#2}}
\endgroup}

\def\brkkk#1#2#3{\begingroup
\tthdump{\hbox to\textwidth{#1}}
\tthdump{{#2}}
\tthdump{\parindent=0pt\newpage{#3}}
\endgroup}
\def\brkkkfirst#1#2#3{\begingroup
\tthdump{\hbox to\textwidth{\hspace{30pt}#1}}
\tthdump{{#2}}
\tthdump{\parindent=0pt\newpage{#3}}
\endgroup}
\def\brkkkf#1#2#3{\begingroup
\tthdump{\hbox to\textwidth{#1} {#2} \newpage\parindent=0pt{#3}}
\endgroup}
\iftth 

\else 
\fi 
\let\bm\bem

\def\bfr{\begin{flushright}}
\def\efr{\end{flushright}}
\iftth 
\def\bn#1{\vspace{}\vspace{}\vspace{}\beq{#1}\lef\vspace{}\vspace{}\vspace{}}
\else 
\def\bn#1{\[{#1}\]}
\fi 
\iftth 
\def\see#1{\hspace{-3pt}
\setcounter{equationp}{\value{equation}}
\addtocounter{equation}{#1}
\setcounter{equationppp}{\value{equationpp}}
\addtocounter{equationpp}{#1}
.qgt.\arabic{section}.{#1}.eqp=\arabic{equationp}.\arabic{equationppp}.cas.)./bf.
\setcounter{equation}{\value{equationp}}
\setcounter{equationpp}{\value{equationppp}}
\hspace{-6pt}}

\else 
\def\see#1{\hspace{-4pt}\addtocounter{equationpp}{#1}
(\ref{\theequationpp})
\addtocounter{equationpp}{-#1}
\hspace{-6pt}$\!$}

\fi 
\def\eeq{\end{eqnarray}}
\iftth
\newcounter{equationp}[section]
\newcounter{tagc}[section] 
\newcounter{equationpp} 
\newcounter{equationppp}
\renewcommand \theequation{\arabic{section}.\arabic{equation}}

\renewcommand \theequationpp{\arabic{equationpp}}

\fi 
\iftth
\def\lefa#1{\end{eqnarray}\eqcard{#1}}
\else 
\def\lefa#1{\label{\theequationpp}\end{eqnarray}\eqcard{#1}}
\fi 
\iftth
\def\lefb{\end{eqnarray}}
\else 
\def\lefb{\label{\theequationpp}\end{eqnarray}}
\fi 
\def\eqcard#1{\vspace{#1}\begin{card}\def\thecard{\theequation}\ccap{}\tthdump{\label{\theequation}}
\end{card}}
\def\eqcards{\begin{card}\def\thecard{\theequation}\ccap{}\tthdump{\label{\theequation}}\end{card}}
\def\tagcard#1#2{
\begin{card}\def\thecard{#2}\ccap{}\tthdump{\label{#1}}\end{card}
\noindent}
\iftth 
\def\tagn#1#2{\addtocounter{tagc}{1}
.oa.{#2}.qgt.
.oa.{#1}.qgt.
.oa.\arabic{section}.tag\arabic{tagc}.qgt.
.oa.\Roman{appendico}.tag\arabic{tagc}.qgt.
}
\else 
\def\tagn#1#2{\vsp{-1.3cm}\addtocounter{tagc}{1}\tagcard{#1}{#2}}
\fi 
\iftth 
\def\lee#1{\end{eqnarray}~.oa.\arabic{section}.\arabic{equation}.qgt.
.oa.equation.\arabic{section}.\arabic{equation}.qgt.
.oa.{#1}.qgt.} 
\else 
\def\lee#1{\label{#1}\label{\theequationpp}\end{eqnarray}\vspace{-1.85cm}\eqcards}
\fi 
\def\lef#1{\nonumber\end{eqnarray}}
\def\lefann{\nonumber\end{eqnarray}\vspace{-0.4cm}\hspace{-5pt}} 
\def\beq{\begin{eqnarray}}
\iftth 
\def\bee#1#2{\addtocounter{equation}{1}\addtocounter{equationpp}{1}~.oa.\arabic{section}.\arabic{equation}.qgt.
.oa.\Roman{section}.\arabic{equation}.qgt.
.oa.\arabic{equationpp}.qgt. 
.oa.equation.\arabic{section}.\arabic{equation}.qgt.
\addtocounter{equation}{-1}.oa.{#1}.qgt.
.bed.
${#2}$ 
.ed1.
\addtocounter{equation}{1}(\arabic{section}.\arabic{equation})
\addtocounter{equationp}{1}\addtocounter{equationppp}{1} 
.ed2.} 
\def\beec#1#2#3{\addtocounter{equation}{1}\addtocounter{equationpp}{1}~.oa.\arabic{section}.\arabic{equation}.qgt.
.oa.\Roman{section}.\arabic{equation}.qgt.
.oa.\arabic{equationpp}.qgt. 
.oa.equation.\arabic{section}.\arabic{equation}.qgt.
\addtocounter{equation}{-1}.oa.{#1}.qgt.
.bedc.
${#2}$
.ed1.
.beed.
${#3}$ 
.ed1.
\addtocounter{equation}{1}(\arabic{section}.\arabic{equation})
\addtocounter{equationp}{1}\addtocounter{equationppp}{1} 
.ed2.} 
\else
\def\bee#1#2{\vspace{0.0cm}\beq{#2}\addtocounter{equationp}{1}\addtocounter{equationpp}{1}\lee{#1}}
\def\beec#1#2#3{
\tthdump{\beq{\hspace{1.2 cm}#2}}
\tthdump{\nn\eeq\vspace{-0.4 cm}}
\tthdump{\beq{#3}}
\addtocounter{equationp}{1}\addtocounter{equationpp}{1}\lee{#1}}

\fi 
  \let\beeplusone\beec

\let\beeplusthree\beeccc
 \let\beeplusfour\beecccc

\def\mathe#1{\begingroup\parindent=0pt\ifmmode{#1}\else${#1}$\fi\endgroup}
\def\@empty{}

\def\nulin#1{\parindent=0pt\parskip=0pt{\tt #1}\hspace{0pt}\parindent=30pt}
\def\numat#1{\begingroup\parskip=1pt{\bf #1} \hspace{0pt}\endgroup\parindent=30pt}

\def\m#1{\ifmmode\mbox{#1}\else ${#1}$\fi}
\iftth 
\def\mbm#1{\ifx #10\newline
\else 
{#1}
\fi}  
\else 
\def\mbm#1{\ifx #10\newline 
\else 
\ifmmode\mbox{#1}\else\numat{#1}\fi
\fi} 
\def\emm#1{\ifmmode\mbox{#1}\else{\em #1}\fi}
\def\embm#1{\ifmmode\mbox{#1}\else{\em #1}\fi}
\fi

\def\link#1#2{
\iftth 
.ataref.{#1}.stcol..foff..bf.{#2}./bf../fon..cas.
\else 
\begingroup\htmladdnormallink{#2}{#1}\endgroup
\fi}
\iftth 
  \def\npa#1#2#3#4{\link{#4}{Nucl. Phys. {\bf A#1}, #2 (#3)}}

  \def\nse#1#2#3#4{\link{#4}{Nucl. Sci. Eng. {\bf  #1}, #2 (#3)}}

  \def\plb#1#2#3#4{\link{#4}{Phys. Lett. {\bf B#1}, #2 (#3)}}
  \def\prl#1#2#3#4{\link{#4}{Phys. Rev. Lett. {\bf  #1}, #2 (#3)}}

   \def\pr#1#2#3#4{\link{#4}{Phys. Rev. {\bf  #1}, #2 (#3)}}

\else 
  \def\npa#1#2#3#4{\link{#4}{Nucl. Phys. {\bf A#1}, #2 (#3)}}

  \def\nse#1#2#3#4{\link{#4}{Nucl. Sci. Eng. {\bf  #1}, #2 (#3)}}

  \def\plb#1#2#3#4{\link{#4}{Phys. Lett. {\bf B#1}, #2 (#3)}}
  \def\prl#1#2#3#4{\link{#4}{Phys. Rev. Lett. {\bf  #1}, #2 (#3)}}

   \def\pr#1#2#3#4{\link{#4}{Phys. Rev. {\bf  #1}, #2 (#3)}}

\fi

\iftth  

\else 

\fi
\def\comment#1#2#3#4#5{
\iftth
\obc\label{comment-#1}
\ifx #20
.ataref.{\$pwd/000ref/\$bwd.htm\#comment-#1}.stcol..foff..bf.\\{--------------------comment-/#1/ {#4} -------------------}\\./bf../fon..cas.\\{\em #5}
\\.ataref.{\$pwd/000ref/\$bwd.htm\#comment-#1}.stcol..foff..bf.\\{----------------------------------------------------------------------------------------------------------------------------------------------------}\\./bf../fon..cas.\\
\else 
.ataref.{\$pwd/#3\#comment-#1}.stcol..foff..bf.\\{--------------------comment-/#1/ {#4} -------------------}\\./bf../fon..cas.\\{\em #5}
\\.ataref.{\$pwd/#3\#comment-#1}.stcol..foff..bf.\\{----------------------------------------------------------------------------------------------------------------------------------------------------}\\./bf../fon..cas.\\
\fi 
\oec 
\else
\ifx #20
\mbm{\obc\link{000ref/$bwd.htm\#comment-#1}{/#1/#4}\oec}
\else 
\mbm{\obc\link{#3\#comment-#1}{/#1/#4}\oec}
\fi  
\fi}

\def\jumponeline{\numat{\vspace{0.40cm}}}
\let\newline=\jumponeline
\iftth 
 
\let\noindent={.noindent.}
\else 

\fi
\let\tta=\mbm
\def\bvb{\begin{verbatim}}
\def\vbm{\begin{verbatim}}
  
\def\tcot#1#2#3{
\if #1c\if #2a{\em{#1#2#3}}\fi\fi
}
\newcommand{\tcom}[1]
{
\ifthenelse{\equal{#1}{true}}{TRUE}{}
\ifthenelse{\equal{#1}{false}}{FALSE}{}
}
\def\etal{$et.al.$}
\def\bfl{\begin{flushleft}}

\def\efl{\end{flushleft}}
 
\def\qq{\mathe{\qquad}}

\def\nn{\mathe{\qquad\qquad\nonumber}}
\tthdump{}
\iftth
\def\pm{¤¤0177;}
\else 
\mathchardef\pmaux="2206
\def\pm{\mathe{\pmaux}}
\fi

\def\sf{\;ba\;}

   \def\abs#1{\ifmmode|{#1}|\else$|$#1$|$\fi}
\def\absu#1#2{\ifmmode|{#1}|^{#2}\else$|#1|^{#2}$\fi}
\def\abss#1#2{\ifmmode|{#1}|_{#2}\else$|#1|_{#2}$\fi}

                  \def\2m#1#2{{$#1{#2}$}}
                \def\3m#1#2#3{{$#1{#2#3}$}}
              \def\4m#1#2#3#4{{$#1{#2#3#4}$}}
            \def\5m#1#2#3#4#5{{$#1{#2#3#4#5}$}}
          \def\6m#1#2#3#4#5#6{{$#1{#2#3#4#5#6}$}}
        \def\7m#1#2#3#4#5#6#7{{$#1{#2#3#4#5#6#7}$}}
      \def\8m#1#2#3#4#5#6#7#8{{$#1{#2#3#4#5#6#7#8}$}}
    \def\9m#1#2#3#4#5#6#7#8#9{{$#1{#2#3#4#5#6#7#8#9}$}}

      \def\brac#1{\mathe{\left({#1}\right)}}
        \def\of#1{\mathe{\left(#1\right)}}

   \def\bracs#1#2{\mathe{\left(#1\right)_{#2}}}

   \def\brak#1{\mathe{\left[#1\right]}}

   \def\brar#1{\mathe{\left\{#1\right\}}}
\def\brars#1#2{\mathe{\left\{#1\right\}_{#2}}}

   \def\bla#1#2{\mathe{\left#1{#2}\right.}}
   \def\bra#1#2{\mathe{\left.{#2}\right#1}}

\def\dirackp{\hspace{1pt}}
\def\diracbp{}
         \def\ket#1{\mathe{\bla{|\!\dirackp}{\bra{>}{#1}}}}
        \def\dket#1{\mathe{\bla{|\!\dirackp}{\bra{>}{#1}}}}

        \def\dbra#1{\mathe{\bla{<}{\bra{\diracbp|}{#1}}}}

       \def\dkb#1#2{\mathe{{\bra{>}{|{#1}}}\!\bla{<}{{#2}|}}}
\let\ketbra\dkb
\let\dagaux=\dagger
           \def\dag{\mathe{\dagaux}}
\let\timesaux=\times
   \def\times{\mathe{\timesaux}} 
\mathchardef\staraux="213F
        
\mathchardef\circaux="220E
        
\def\vecaux{\mathaccent"017E }
         \def\vec#1{\mathe{\vecaux{#1}}} 
     
\def\widetildeau{\mathaccent"0365 }

\def\widetilde#1{\mathe{\widetildeau{#1}}}

 \def\dbokt#1#2#3#4{\mathe{\bla{<}{#1}|\dirackp{#2}|\bra{>}{#3}\!\!\bla{<}{#3}|\dirackp{#4}|\bra{>}{#1}}}
 
    \def\dbok#1#2#3{\mathe{\bla{<}{#1}|\dirackp{#2}|\bra{>}{#3}}}
       
        \def\dbko#1{\mathe{\left<{#1}\right>}}
         
     \def\dbkos#1#2{\mathe{\left<{#1}\right>_{#2}}}

       \def\rbk#1#2{\mathe{{(}{{#1}|}{{#2}{)}}}}

\let\braket\ddbraket

\def\lim#1{\mathe{\mathop{\tt lim}\limits_{#1}}}
\iftth                  

\def\cos{.hspac.{\tt cos}}

\else 

\def\cos{\mathe{{\mbox{\;cos}}}}

\fi 

\let\over=\choose 
\def\jover{\atopwithdelims..}

\iftth

     \def\leftarrow{{\lz ¤¤8592;}}
    \def\rightarrow{{\lz ¤¤8594;}} 

\else 

\def\leftrightarrowaux{\leftarrow\!\!\!\rightarrow}

\let\rightleftarrow\leftrightarrow 
\let\logaux=\log
\def\log{\mathe{\logaux}}
\let\leftarrowaux=\leftarrow
\let\rightarrowaux=\rightarrow 

     \def\leftarrow{\mathe{\leftarrowaux}}
    \def\rightarrow{\mathe{\rightarrowaux}}

\fi 

      \def\da{\mathe{d}}
      
      \def\fa{\mathe{f}}
      \def\ga{\mathe{{\tt g}}}
      \def\ha{\mathe{h}}
      \def\ia{\mathe{i}}
      \def\ja{\mathe{{\tt j}}}
      \def\ka{\mathe{k}}
      \def\la{\mathe{l}}

      \def\na{\mathe{n}}
      
      \def\pa{\mathe{p}}
      
      \def\ra{\mathe{r}}
      \def\sa{\mathe{s}}
      \def\sl{\mathe{s}}
      \def\ta{\mathe{t}}
      \def\ua{\mathe{u}}

      \def\xa{\mathe{x}}
      \def\ya{\mathe{y}}
      
       \def\ac{\mathe{A}}
       
       \def\cc{\mathe{C}}

       \def\fc{\mathe{F}}

       \def\mc{\mathe{M}}
       \def\nc{\mathe{N}}
       \def\oc{\mathe{O}}
       \def\pc{\mathe{P}}

       \def\sc{\mathe{S}}
       \def\su{\mathe{S}}
       
       \def\uc{\mathe{U}}
       \def\vc{\mathe{V}}
       
       \def\xc{\mathe{X}}

\iftth 
\else
 \def\min{\mathe{\tt{min}}} 
 \def\max{\mathe{\tt{max}}} 
 \def\cut{\mathe{\tt{cut}}}

\def\initial{\mathe{\tt{initial}}}
\def\final{\mathe{\tt{final}}} 
 
\fi

              \def\cdotaux{\mathinner{\cdotp\cdotp\cdotp}}
\def\cdots{\mathe{\cdotaux}} 

\mathchardef\sumaux="1350
       \def\sum{\mathe{\sumaux}}
 \def\sigsu#1#2{\mathe{\sumaux_{{#1}}^{#2}}}

    \def\sigsil#1{\mathe{\sumaux_{{#1}}}}
\def\sigss#1#2{\mathe{\sumaux_{{#1}\jover{#2}}}}

\def\sigssu#1#2#3{\mathe{\sumaux_{{#1}\jover{#2}}^{#3}}}
 \def\sumsu#1#2{\mathe{\sumaux_{({#1})}^{#2}}}
      \def\sums#1{\mathe{\sumaux_{({#1})}}}

   \def\sumsne#1{\mathe{\sumaux_{({#1})}}}

\def\sumss#1#2{\mathe{\sumaux_{{#1}\over{#2}}}}
\def\sumst#1#2{\mathe{\sumaux_{({#1})({#2})}}}

\let\intaux=\int 
          \def\int{\mathe{\intaux}}

    \def\intsu#1#2{\mathe{\intaux_{{#1}}^{#2}}} 
\let\ointaux=\oint 
          
        \def\doint{\mathe{\ointaux\ointaux}}

\mathchardef\inftaux="0231
\def\infty{\mathe{\inftaux}}
\def\prodsu#1#2{\mathe{\prod_{#1}^{#2}}}

   \def\prods#1{\mathe{\prod_{#1}}}
   
               \let\inaux=\in
    \def\in{\mathe{\,\inaux}\,}
   \def\bel{\mathe{\,\inaux\,}}

\let\neqaux=\neq

\def\dif{\mathe{\neqaux}}

    \def\as#1{\mathe{a_{#1}}}
    
 \def\asu#1#2{\mathe{a_{#1}^{#2}}}

 \def\aus#1#2{\mathe{a^{#1}_{#2}}}
   \def\aas#1{\mathe{a_{#1}}}

    \def\bs#1{\mathe{b_{#1}}}

 \def\bus#1#2{\mathe{b^{#1}_{#2}}}

   \def\cas#1{\mathe{c_{#1}}}

\def\dasu#1#2{\mathe{d_{#1}^{#2}}}

   \def\das#1{\mathe{d_{#1}}}
   
    \def\eu#1{\mathe{{\tt e}^{#1}}}
   \def\eub#1{\mathe{{\tt e}^{(#1)}}}

   \def\fas#1{\mathe{f_{#1}}}

\def\hasu#1#2{\mathe{h_{#1}^{#2}}}

   \def\ias#1{\mathe{{\tt i}_{#1}}}

   \def\jas#1{\mathe{{\tt j}_{#1}}}

    \def\ks#1{\mathe{{\tt k}_{#1}}}

   \def\kas#1{\mathe{{\tt k}_{#1}}}

    \def\ls#1{\mathe{{\tt l}_{#1}}}

    \def\ms#1{\mathe{{\tt{m}_{#1}}}}
  
   \def\mas#1{\mathe{\tt{m}_{#1}}}

    \def\ns#1{\mathe{n_{#1}}}

   \def\nas#1{\mathe{n_{#1}}}

    \def\pp#1{\mathe{p^{#1}}}

    \def\rs#1{\mathe{r_{#1}}}

   \def\ras#1{\mathe{r_{#1}}}
   
   \def\up#1{\mathe{^{#1}}}
   \def\sub#1{\mathe{_{#1}}}

    \def\ss#1{\mathe{s_{#1}}}

   \def\sas#1{\mathe{s_{#1}}}

    \def\us#1{\mathe{u_{#1}}}

    \def\xs#1{\mathe{x_{#1}}}
    \def\xu#1{\mathe{x^{#1}}}

   \def\xas#1{\mathe{x_{#1}}}
   \def\xau#1{\mathe{x^{#1}}}

    \def\yu#1{\mathe{y^{#1}}}

   \def\yau#1{\mathe{y^{#1}}}

      \def\hp{\mathe{h'}}

      \def\pp{\mathe{p'}}

      \def\xp{\mathe{x'}}
      \def\yp{\mathe{y'}}

     \def\cap{\mathe{c'}}

     \def\hap{\mathe{h'}}

     \def\nap{\mathe{n'}}
     
     \def\pap{\mathe{p'}}

   \def\xps#1{\mathe{x'_{#1}}}

   \def\xpu#1{\mathe{x'^{#1}}}

   \def\ypu#1{\mathe{y'^{#1}}}

     \def\asl{\mathe{\as{l}}}
    \def\asnu{\mathe{\as{\nu}}}
    
   \def\auspl{\mathe{a^{\dagger}_{l}}}

  \def\auspnu{\mathe{\as{\nu}^{\dagger}}}
  
    \def\bsmu{\mathe{\bs{\mu}}}

  \def\buspmu{\mathe{\bs{\mu}^{\dagger}}}

    \def\xspl{\mathe{x_{pl}}}
   \def\xspnu{\mathe{x_{p\nu}}}
     
   \def\xshmu{\mathe{x_{h\mu}}}

    \def\ucsk{\mathe{U_{k}}}
    \def\fcsh{\mathe{F_{h}}}
    
    \def\fcsp{\mathe{F_{p}}}
   \def\fcpsh{\mathe{F'_{h}}}
   
   \def\fcpsp{\mathe{F'_{p}}}
   \def\epsi{\mathe{\epsilaux_{i}}}

   \def\bcs#1{\mathe{B_{#1}}}

   \def\fcs#1{\mathe{F_{#1}}}

   \def\mcs#1{\mathe{M_{#1}}}
   \def\mcu#1{\mathe{M^{#1}}}
   \def\ncs#1{\mathe{N_{#1}}}

   \def\ocu#1{\mathe{O^{#1}}}
   \def\pcs#1{\mathe{P_{#1}}}

   \def\scs#1{\mathe{S_{#1}}}

   \def\ucs#1{\mathe{U_{#1}}}
   
   \def\vcs#1{\mathe{V_{#1}}}

\def\vcus#1#2{\mathe{V^{#1}_{#2}}}
\def\vcsu#1#2{\mathe{V_{#1}^{#2}}}

      \def\mcp{\mathe{M'}}

      \def\ucp{\mathe{U'}}

   \def\fcps#1{\mathe{F'_{#1}}}

   \def\fcps#1{\mathe{F'_{#1}}}

\let\sqrtaux=\sqrt
\def\sqrt#1{\mathe{\sqrtaux{#1}}}
\iftth 
\else
                       \let\hbaux=\hbar
          \def\hbar{\mathe{\hbaux\hspace{1pt}}}

\fi
\let\lamba=\lambdabar
     \def\lambdabar{\mathe{\lamba}}
        \let\muaux=\mu
\mathchardef\nuaux="0117
         
                  \iftth 
                  \else 

\mathchardef\alphaux="010B
\mathchardef\betaux="010C
\mathchardef\gammaux="010D
\mathchardef\deltaux="010E
\mathchardef\Deltaux="7001
\mathchardef\epsilaux="010F
\mathchardef\etaux="0111
\mathchardef\tauaux="011C
\mathchardef\thetaux="0112
\let\alpha\undefined
\let\beta\undefined
\let\gamma\undefined
\let\delta\undefined
\let\Delta\undefined
\let\epsilon\undefined
\def\alpha{\mathe{\alphaux}}
\def\beta{\mathe{\betaux}}
\def\gamma{\mathe{\gammaux}}
\def\delta{\mathe{\deltaux}}
\def\Delta{\mathe{\Deltaux}}
\def\epsilon{\mathe{\epsilaux}}
\def\eta{\mathe{\etaux}}
\mathchardef\ggaux="321D

\mathchardef\llaux="321C

\mathchardef\phiaux="011E
\mathchardef\Phiaux="7008
 \def\phi{\mathe{\phiaux}}
 \def\Phi{\mathe{\Phiaux}}

\mathchardef\xiaux="0118
\mathchardef\chiaux="011F
\mathchardef\Xiaux="7004
\mathchardef\geaux="3215 
\mathchardef\rhoaux="011A
\mathchardef\psaux="0120
\mathchardef\Psaux="7009
\mathchardef\omegaux="0121
\mathchardef\Omegaux="700A
\let\otimx=\otimes
       \def\otimes{\mathe{\otimx}}
\mathchardef\propaux="322F

        \let\propto=\prop
\mathchardef\approxaux="3219
        \def\approx{\mathe{\;\approxaux\;}}
        \let\app=\approx
         \let\lambaux=\lambda
      \def\lam{\mathe{\lambaux}}
     
   \def\lambda{\mathe{\lambaux}}

   \def\lams#1{\mathe{\lambaux_{#1}}}
  \def\lambs#1{\mathe{\lambaux_{#1}}}

\def\lamsu#1#2{\mathe{\lambaux_{#1}^{#2}}}

         \let\Lambaux=\Lambda

   \def\Lambda{\mathe{\Lambaux}}

    \def\bes#1{\mathe{\betaux_{#1}}}

      \def\bep{\mathe{\betaux'}}
      
     \def\gamp{\mathe{\gammaux'}}

    \def\des#1{\mathe{\deltaux_{#1}}}

    \def\eps#1{\mathe{\epsilaux_{#1}}}

\mathchardef\leaux="3214 
\let\gtaux=>
\let\ltaux=<
\def\gt{\mathe{\gtaux}}

\def\le{\mathe{\;\leaux\;}}

\def\a{\mathe{\alphaux}}
\def\b{\mathe{\betaux}}
\def\g{\mathe{\gammaux}}
\def\d{\mathe{\deltaux}}
\def\e{\mathe{\epsilaux}}

\def\n{\mathe{\nuaux}}
\def\rho{\mathe{\rhoaux}}

\def\tau{\mathe{\tauaux}}
\def\t{\mathe{\tauaux}}

\def\theta{\mathe{\thetaux}}
\def\al{\mathe{\alphaux}}
\def\be{\mathe{\betaux}}
\def\de{\mathe{\deltaux}}
\def\ep{\mathe{\epsilaux}}
\def\mu{\mathe{\muaux}}

\def\nu{\mathe{\nuaux}}

   \def\nus#1{\mathe{\nuaux_{#1}}}

    \def\nusi{\mathe{\nuaux_{i}}}
\def\om{\mathe{\omegaux}}
\def\ome{\mathe{\omegaux}}
        
    \def\Omega{\mathe{\Omegaux}} 
\let\piaux=\pi
\let\piauxc=\Pi

\def\pi{\mathe{\piaux}}
\def\Pi{\mathe{\piauxc}}

\def\psi{\mathe{\psaux}}
\mathchardef\kappaaux="0114

\def\kappa{\mathe{\kappaaux}} 
\def\xi{\mathe{\xiaux}}
\def\Xi{\mathe{\Xiaux}}

\def\chi{\mathe{\chiaux}}

    \def\oms#1{\mathe{\omega_{#1}}}
    \def\omu#1{\mathe{\omega^{#1}}}

   \def\psis#1{\mathe{\psaux_{#1}}}

     \def\delc{\mathe{\Delta}} 
    
   \def\deltac{\mathe{\Delta}}

   \def\delcs#1{\mathe{\Delta_{#1}}}

     \def\lamc{\mathe{\Lambda}}

                            \fi 

\def\eq{\mathe{=}}

\def\mi{\mathe{-}}
\def\pl{\mathe{+}}
 
\iftth
\mathchardef\capux="225C
\mathchardef\cap="225C
\mathchardef\intersec="225C
\mathchardef\intersection="225C
\mathchardef\cupux="225B
\mathchardef\cup="225B
\mathchardef\union="225B
\let\fracux=\mathe{\frac}
\else 
\mathchardef\capux="225C
\def\cap{\mathe{\capux}}
\let\intersec=\cap
\let\intersection=\cap
\let\fracux=\frac
\mathchardef\cupux="225B
\def\cup{\mathe{\cupux}}
\def\frac#1#2{\mathe{\fracux{#1}{#2}}}
\fi

\let\union=\cup
\iftth
\def\fracs#1#2{
\ifmmode\frac{#1}{#2}\else\mathe{{#1}\left/{#2}\right.}\fi} 
\else 
\def\fracs#1#2{
\ifmmode\fracux{#1}{#2}\else\mathe{\hspace{-3pt}{#1}\hspace{-4pt}\left/{#2}\right.}\fi} 
\fi 
 
\def\cala{\mathe{\begingroup \cal{A}\endgroup}}

\def\cald{\mathe{\begingroup \cal{D}\endgroup}}

\def\call{\mathe{\begingroup \cal{L}\endgroup}}
\def\caln{\mathe{\begingroup \cal{N}\endgroup}}

\def\calds#1{\mathe{{\begingroup \cal{D}\endgroup}_{#1}}}

\def\calns#1{\mathe{{\begingroup \cal{N}\endgroup}_{#1}}}

\def\calcsj{\mathe{{\begingroup \cal{C}\endgroup}_{j}}} 
\def\calwsj{\mathe{{\begingroup \cal{W}\endgroup}_{j}}} 
\def\caldsk{\mathe{{\begingroup \cal{D}\endgroup}_{k}}} 
\def\calnsk{\mathe{{\begingroup \cal{N}\endgroup}_{k}}} 
 
\def\lapt#1{\call\brar{{#1}}}

\iftth

\else

 \def\fe#1{\begingroup\small\m{^{#1}}\endgroup Fe}

\fi

\tthdump{   }
\tthdump{   }
\tthdump{   }
\tthdump{   }
            
\tthdump{   }
\tthdump{   }
\tthdump{   }
\tthdump{   }
             \def\tn{{\bf TNG}}
            \def\tng{{\bf TNG}}

            \def\exm{EXM}
            \def\EXM{EXM}
            \def\oxm{EXM}
            \def\OXM{EXM}
            \def\DMF{DMF}
            \def\DFM{DMF}

             \def\tran{{\bf TRANSNU}}
             \def\trans{{\bf TRANSNU}}

            \def\ripl-ii{RIPL-2}
            \def\ripl{RIPL-2}
            \def\ripl2{RIPL-2}

            \def\ripl-iis{{RIPL-2} }
            
            \def\ripl2s{{RIPL-2} }

            \def\ripl-iib{{RIPL-2} }
            
            \def\ripl2b{{RIPL-2} }

 \def\pe{PE}
 \def\cn{CN}
 \def\qe{QE}

\def\cns{CN }

\def\capit#1{
\if #1=a A\fi
\if #1=b B\fi
\if #1=c C\fi
\if #1=d D\fi
\if #1=e E\fi
\if #1=f F\fi
\if #1=g G\fi
\if #1=h H\fi
\if #1=i I\fi
\if #1=k K\fi
\if #1=j J\fi
\if #1=l L\fi
\if #1=m M\fi
\if #1=n N\fi
\if #1=o O\fi
\if #1=p P\fi 
\if #1=q Q\fi
\if #1=r R\fi
\if #1=s S\fi
\if #1=t T\fi
\if #1=u U\fi
\if #1=v V\fi
\if #1=w W\fi
\if #1=x X\fi
\if #1=y Y\fi
\if #1=z Z\fi
} 

\tthdump{}
\tthdump{}
\tthdump{}
\tthdump{}
\tthdump{}
\tthdump{}
\tthdump{}
\tthdump{}
\tthdump{}
\tthdump{}
\tthdump{}
\tthdump{}

\def\smod{Shell Model}

\begin{document}

  \pagestyle{myheadings}
  \markright{\thepage}


\vspace{-3.5cm} 

\title{Application of the Direct Microscopic Formalism to Nuclear Data Evaluation}

\author{F. B.  Guimaraes}

\vspace{-1.7cm} 
\address{
Instituto de Estudos Avan\c cados/DCTA,\\ 
12228-001 S\~ao Jos\'e dos Campos, S\~ao Paulo, Brazil \\ 
e-mail: fbraga@ieav.cta.br} 

\vspace{-1.0cm} 
\begin{abstract}
\mz{\it{We review the basic definitions of the direct microscopic formalism  (DMF) and
the corresponding model code TRANSNU, to describe pre-equilibrium nuclear systems, as
elements of the grand canonical ensemble. 
We analyze its inconsistencies, especially the impossibility of exact solution of the
nuclear many-body problem, and propose solutions. 
   
We use the strong dependence of pre-equilibrium emissions on the ratio of transition
rates, in code \tng, to propose a redefinition of the parameters in the master equation,
and obtain smoother excitation functions in the energy regions where different exciton
classes contribute.  We compare the transition rates of TRANSNU with the phenomenological
ones for the estimation the excitation function of a few \pa-induced reactions on
\fe{56}, and obtain reasonable descriptions for activation energy, local maxima and
average magnitude.

Despite the inconsistencies and a few remaining numerical problems, related with
non meaningful noises in the strong oscillations of the TST, especially with the
excitation energy and for low exciton number, we find the results of the presesnt work
very promising, indicating that the \DFM\ can be used as a more precise and physically
meaningful approach for the study of nuclear systems in pre-equilibrium than the
traditional statistical modesls. }}
\end{abstract} 

\secto{Introduction}

The analysis of the pre-equilibrium phase of nuclear reactions (\pe) is usually made by
considering more or less approximate phenomenological formalisms, in which the
microscopic description is treated semi-classically, and statistical approximations are
used to describe the parameters of the system.

One of the most common approaches is to consider the general \embm{exciton} concept
originally proposed by Griffin\cite{g66}, to analyze nuclear states and particle
emissions before the formation of the compound nucleus. This approach  can be 
divided into two main semi-classical lines:\cite{semiclasshm} the \embm{hybrid model} of
Blann and Vonach and the \embm{Standard Exciton Model}
(\oxm)\cite{semiclasssem,k72,b68,w71,e60}. As explained in Refs.\cite{pc,bp86} both
approaches have the same basic description of the evolution of the nuclear system at the
\pe\ stage, adopting the idea of a chain of increasingly complex systems generated by the
residual interactions, in which the initial nuclear energy is more and more evenly shared
by all excitable nuclear degrees of freedom.

 The essential difference between these two semi-classical approaches is the opposite
 assumption about the level of configuration mixing of the evolving system at each step
 of the chain: the hybrid model assumes ``no mixing" between the various ``classes of
 excitons", each class defining a level of ``complexity", while the \oxm\ assumes
 ``perfect mixing" between intra-class transitions, with or without particle emissions,
 i. e., that the energy is perfectly shared among the excitons at each step of the
 complexity chain and the occupation probabilties of all configurations or each class are
 the same.

In these models
the nuclear ``complexity" is then defined by the number of excited \embm{single particle
states} (sp-states) in comparison with the fundamental state, where the total nuclear
excitation is zero and all component nucleons occupy the lowest possible energies levels
up to a maximum called Fermi level, \eps{\fc}.

To describe the excited nuclear system one initially considers the fundamental state and
the sp-states that can be possibly excited on it, and call these sp-states ``holes" when
created below \eps{\fc} or ``particles" when created above it.  In addition, the
\embm{initial sp-states above} \eps{\fc} are also defined as holes.

All these sets of ``excitable" states are model dependent
and, in particular, the number of holes above \eps{\fc} can be arbitrarily large.
If at a given moment of the evolution of the nuclear system there  are ``\ha" holes ``\pa" particles
the total number
\bee{exm}{\na=\pa+\ha\ \; }
of excited sp-states is called the \embm{exciton number} of the system. For nuclear
states with excitation energy greater than zero, one assumes that new particles can be created only in
states previously occupied by holes and, vice-versa, holes can be created only in states previously
occupied by particles, corresponding to the ``phyical concept" of excitons.

In this work we follow the usual definition of Model Space\cite{fbgarXiv1,obp,w69,w71,b68}
and expand the above concept to consider the more abstract or ``mathematical concept" of
particles and holes as two \embm{independent fermion fields}, i. e., that can be created
or destroyed independently, to permit a simpler algebraic definition of nuclear excited
states and then add the physical constraint that a ``particle" and a ``hole" cannot be in
the same state at the same time.

Therefore, we define independent sets of levels for the particle and hole fields and
\eps{\fc} also as an independent phenomenological parameter of the model,
related to the \embm{depth} of the long range attracting nuclear potential.
The interaction of the excited sp-states is usually defined by a phenomenological
``residual potential", or ``mean \embm{field"} added to the free part of the nuclear
Hamiltonian, and the basis of sp-states can be defined self-consistently or
by an \embm{ad hoc}
phenomenological function.

The resulting nuclear system is then, by definition, in a state of
\embm{quasi-equilibrium} (\qe), i.  e., its state changes can be described by first order
perturbations over the free movement description, considered as in an ``approximate state
of equilibrium". The latter is defined essentially by the mean-field itself, where a
non disturbed equilibrium corresponds to zero mean-field.

To simplify the present analysis and permit a direct comparison with semi-classical
studies, the basis adopted for the single particle wave-functions in this work is either
the phenomenological Hamonic Oscillator basis (H.O.) or the approximated ``constant
spacing" basis, in which the spin-orbit splitting of the eigen-levels
of the H.O.  Hamiltonian is neglected, but the other quantum numbers and the
eigenfunctions themselves of the H.O. basis are kept,\cite{bg77} with an arbitrary
constant interspacing between single particle states.

The connection with actual nuclear systems at the \pe-stage is realized by assuming
that the sp-energies cannot be greater than a given phenomenological maximum, which
coincides with the maximum of the energy of the ``hole" sp-states at the initial time,
and that particles and holes can only be created or destroyed simultaneously, i. e., in
pairs of particle-hole sp-states (ph-pairs).  Then, the total number of excitons can only
vary in steps of \deltac\na=\pm2 or be constant.

From a given initial excited state, for example by the absorption of an incident
nucleon on a given target nucleus, the system is supposed to evolve by increasingly
sharing the total excitation among the largest possible number of sp-states, increasing
the number of excitons.

The sharing occurs at the microscopic level as a consequence of the residual interactions
between sp-states and the increase of \na\ is counterbalanced by possible PE emission of
particles or annihilation of a ph-pairs. This produces an average systematic increase of
number of excitons up to a maximum that defines the ``most probable exciton number" at
the PE stage, \widetilde{\na}, beyond which the system is supposed to evolve
\embm{preferrably} towards full equilibrium (compound nucleus state) instead of by
emission of more particles at the PE stage.

In the \OXM\ the rates of transition between nuclear states of increasing complexity are
usually given in an essentially phenomenological way.  In particular, the transition
matrix elements, describing the residual mean-field, are considered as phenomenological
\embm{constants} or simple functions of the nuclear excitation energy, \uc, as
exemplified in Ref.\cite{kon04}.

In addition, the \EXM\ relies on approximations of statistical nature that may not be
very precisely defined for nuclear systems at microscopic level, which often makes it
difficult to evaluate the importance of the details of the microscopic interaction in the
description of the PE process. For example, the semi-classical
formulations\cite{semiclasshm,semiclasssem} give the density of states for a given {\uc}
as a convolution of the densities for {\pa} and {\ha},\cite{w71}
\bm{\oms{\pa\ha}\brac{\uc} = \intsu{0}{\uc} \oms{\pa}\brac{\uc}\oms{\ha}\brac{\uc\mi\uc} \da\uc \;, }
which can be deduced from the continuum approximation (\emm{CAP}), but does not result
directly from the quantum microscopic description.

The traditional approach to the Shell Model defines the moments of the Hamiltonian in
terms of Laplace transforms and their inverse to obtain expressions like \see{-0} for the
nuclear density and in the microscopic description of Ref.\cite{obp} a similar approach
is followed to obtain the expressions for the transition strengths.

In this work we use the \embm{direct microscopic formalism} (\DMF), and the corresponding
model code TRANSNU, of Refs.\cite{obp,fbgarXiv1} in which one is able to produce the
results of the Shell Model without effectively relying on the validity of \emm{CAP} for
nuclear levels
and other common approximations of the semi-classical models.

In this respect the \DMF\ is a more natural and appropriate
description of nuclear transition strengths, bringing a complementary
view to the usual statistical description of PE states.

One of the central aspects of the \DMF\ is the proportionality of the
degeneracy of a given nuclear state (with given total excitation energy, \uc, angular
momentum,  \mc, and number of excitons, \m{p} and \m{h}), \da(\uc,\mc,\pa,\ha), to the
corresponding density of nuclear states
\bm{\da(\uc,\mc,\pa,\ha) \propto \om(U,M,p,h) , }
i. e., the quantization of \om(\uc,\mc,\pa,\ha), which permits to connect to the traditional
 approach, using \emm{CAP}, and reinterpret the moments of the nuclear Hamiltonian, in
 this limit, in terms of usual convolutions over the excitation energy of the moments of
 less excited states in accordance with the general proposal of Ref.\cite{obp}.

Throughout this paper we use the two letters ``sp" to designate ``single particle".

In \ref{section.2} we present the basic formal definitions of the direct microscopic
algebra,\cite{obp,fbgarXiv1} the definitions of the Shell Model and of the nuclear
density in connection with the Darwin-Fowler statistics and the computation of the
momenta of the nuclear Hamiltonian. We briefly review the importance of \emm{CAP} and the
grand canonical ensemble of Statistical Mechanics to connect with the Laplace transform,
as it is done in the traditional statistical
pre-equilibrium approaches, and how the \DFM\ provides a more direct
and physically meaningful way to describe the same systems.

In \ref{section.3} we review some aspects of the statistical phenomenological
approaches that apply to the present discussion, especially in connection with
code \tng's definitions and use of transition rates during the pre-equilibrium stage,
inspired by the original approach of Griffin.\cite{g66}

In \ref{section.4} we present the main results of this paper. We
compare numerically the transition strengths of TRANSNU (TST) and transition rates (TR's)
of TNG and another independent \OXM\ model and apply them to the estimate of the
excitation function of a few \pa-induced reactions on \fe{56}.

We use the different designation of ``TST" and ``TR" for the parameters calculated by
TRANSNU and the phenomenological models, respectively,
throughout the paper despite the fact that these parameters are describing the
same physical phenomenon, i. e. the rates of transition between classes of nuclear states
during the \pe\ stage, to reinforce the essentially different ways by which they are
calculated in each case.

In \ref{section.5} we present a brief review of the main results of the paper and our
evaluation of what we were able to obtain with TRANSNU and the \DFM\ in the present work.

\secto{The Direct Microscopic Formalism}

To make this presentation clearer and more self-contained we review in this section the basic
results of the \DMF\ defined in  Ref.\cite{obp,fbgarXiv1}.

The usual statistical description of the nuclear system inspired in the \smod,\cite{b68}
can be defined as a limiting case of the ``equidistant spacing model" as follows. The nuclear
mass number is given by
\bee{det1}{ \ac=\sums{i}\ns{i} \;, }
where \ns{\ia} are the occupation numbers of single particle states (sp-states)
(satisfying: \ns{\ia}\in\brar{0,1}) associated with the corresponding set of energies
\bee{det2}{ \epsilon_i = \nusi \epsilon \;,}
where the \nusi\ are integers and \ep\ is a fixed real number, defining a
constant spacing between any two consecutive sp-levels of the approximate ``equidistant
spacing model", but \ep\ can also be considered as an average spacing of more realistic
bases for the sp-states as e. g. the H.O. basis. The total nuclear energy is then given
by
\bee{det3}{ U = \caln \ep =  \sums{i} n_{i} \nus{i} \ep \;}
and if \ka\ elements of the set \m{\brar{\nusi,i=1,..,\infty}} are
equal, with \ka\gt1, the corresponding elements,
\bee{det4}{\nus{i1}=\cdots=\nus{ik}\bel\brar{\nusi,i=1,..,\infty}\,,}
define the \embm{set of degenerate sp-states} or the \embm{energy level} (sp-level) to
which these states belong
\bee{det41}{\epsi=\nus{i1}\ep=\cdots=\nus{ik}\ep\,.}

A nuclear \embm{configuration} (nuclear state) is defined as a set of bound sp-states of
nucleons, characterized by the set of quantum numbers of each component sp-state. Then,
the number of elements of the set of nuclear configurations associated with a \emm{given
nuclear energy} \uc\ is, by definition, the \embm{degeneracy} of the corresponding
\emm{nuclear level}, \cald(\uc), characterized by \uc.

   For given {\ac} and {\uc}, the nuclear degeneracy is equal to the number of
   solutions of the Eqs.\reg{det1} and \reg{det3}.
Therefore, if one considers the microscopic distribution of nuclear states as a function of the
nuclear energy, \emm{the cummulative number of states increase in steps equal to the
degeneracy of each nuclear level}.

If the sp-states are excitons, as defined at the Introduction, the particle and hole
states may have energies conveniently defined with respect to \eps{\fc} and the nuclear
energy will be equal to the {\em excitation energy}.

One should notice that in the microscopic description the \embm{degeneracy of the sp-levels} is
defined by the characteristic features of the nuclear Hamiltonian, i. e., by the set of
nuclear forces and the symmetries of the nuclear system that define the resulting
potential on each component nucleon and the consequent structure of sp-states.

On the other hand, due to the above definition of nuclear state, the \embm{nuclear
degeneracy} has a ``combinatorial" meaning, in terms of the distribution of nucleons into
the ``pre-defined" structure of sp-states, resulting from the solution of the nuclear
many-body problem.

In this context, if \caln\ is the cumulative number of nuclear levels up to energy \uc,
the nuclear level density in the continuous approximation (CAP) is given
by\cite{fbgarXiv1}
\bee{lev3}{ \rho(\uc)=\frac{\da\caln}{\da\uc} \app \frac{\caldsk(\uc)}{\ep} }
or, using the Darwin-Fowler method, one may consider the generating function for the
corresponding grand canonical ensemble, given by\cite{b68}
\bee{det5}{ f(x,y)= \prods{i}\brac{1+\xa\yau{\nus{i}}}= \prods{i}\brac{1+\xas{i}}
\;,  }
where, \ia\ are the indices for sp-states as in \see{-6} and \see{-4} and \xa\ accounts
only for the \emm{number}  of sp-states in each nuclear configuration, and is the same
parameter for all sp-states. Both \xa\ and \ya\ have physical meaning under the
statistical description of the grand canonical ensemble, but \xa\ has a more strictly
combinatorial meaning while \ya\ is related with the probability distribution associated
with the various microscopic systems of the ensemble.

Then, the nuclear level density can be directly defined, using the Cauchy theorem, as the
value of the of the generating function at an adequate {\em pole} divided by
{\ep},\cite{b68}
\bee{det6}{ \rho(\ac,\uc) = \frac{1}{\brac{2\pi i}^2 \ep}
\doint\frac{f(x,y)\da\xa\da\ya}{\xu{A+1}\yu{\caln + 1}} \;, }
while the generating function can be rewritten as
\bee{det16}{
f(x,y) = 1 + x \sums{j} y^{\nus{J}} + x^2 \sums{j1, j2} y^{\brac{\nus{J1}+\nus{J2}}} + \cdots
+ x^A \sums{j1, \cdots, jA} y^{\brac{\nus{J1}+ \cdots + \nus{JA}}} + \cdots \;.
}
Eq.\see{-0} describes, for example, all nuclear systems with all possible ``mass
numbers" (number of sp-states in each ``microcanonical state" or ``microstate") and energies
(total energy of the microstate).
In other words, each configuration of sp-states is also a microstate of the canonical ensemble
with fixed mass number and temperature,\cite{reif} and the term proportional to \xau{\ac}
is the sum over all possible configurations with fixed nuclear mass \ac\ and variable
energy. These canonical ensembles are the main focus of the \DMF.

Equation \see{-0} can be rewritten as
\beec{det24}{
  f(x,y) = 1 + x   \sums{k1} \calds{k1}(\ucs{k1},1) y^{\calns{k1}(1)} + \nn}{
               x^2 \sums{k2} \calds{k2}(\ucs{k2},2) y^{\calns{k2}(2)} +
      \cdots + x^A \sums{k } \calds{k }(\ucs{k },A) y^{\calns{k }(A)} + \cdots + }
where
for each nuclear level, {\ucsk}, corresponds usually many different configurations of
sp-states and for each {\ac} one has
\bee{det21}{ \sums{j1, \cdots, jA} y^{\brac{\nus{j1} + \cdots + \nus{jA}}}
= \sums{k} \caldsk y^{\calnsk} }
where {\caldsk = \caldsk(\ucsk,\ac)} is the degeneracy of the nuclear level
{\ucsk}, and
\bee{det22}{ \calnsk = \sigsu{i=1}{A} \nus{ki} = \frac{\ucsk}{\ep}\;. }

In \see{-2} different indices, \ka1, \ka2, etc., have been used for each term to
reinforce the idea that the corresponding nuclear levels may not be the same. In these
expressions the sum over \ka\ is equivalent to the sum over \ucs\ka, then \fa(\xa,\ya)
can be rewritten as a sum over nuclear levels,
\bee{det230}{ f(x,y) = \sums{A,\uc} \cald(\ac,\uc)\xau{A}\yau{\uc/\ep} \;,}
and also as a sum over \embm{individual configurations}, with all degeneracies equal one,
\bee{det230}{ f(x,y) = \sums{conf} \xau{A} \yau{\uc/\ep} \;.}

%
\subsecto{Expected values in the \DFM}

We consider here a given nuclear excited state, with variable mass number \ac\ and energy
\uc, belonging to the corresponding grand canonical distribution defined over all nuclear
configurations obtained as the solutions of the Eqs.\ree{det1} and \ree{det3}.

One may always assume that the elements of the set of characteristic integers of the
energies of sp-states, {\brar{\nus{\ia}}} in \ree{det3}, are ordered according to
increasing values as a function of \m{\ia} and, by definition, there are only {\ac}
nonzero elements in the set \brar{\nas{\ia}}.

Then the nuclear state, \dket{\psis{\ac}}, can be represented by a product of \ac\
sp-states, which we assume to have constant energy as a function of the time \ta\ of the
evolution of the system, so that each microstate is described by the independent particle
model (IPM).\cite{ring_shuck,fetter} 

Then, \ket{\psis{\ac}} can be written as a linear combination, Slater
determinant, of the corresponding complete set of eigenvectors of the single particle
Hamiltonian. To simplify the presentation we consider only one term of the determinant to
represent \ket{\psis{\ac} } 
\bee{IPM}{
\dket{\psis{\ac}} =  \prodsu{i=1}{\ac} \dket{\psis{i}(\ta)} =
\prodsu{i=1}{\ac} \sums{\kas{i}} \cas{\ia\kas{i}} \eu{\ia\eps{\kas{i}}t/\hbar}
\dket{\us{\kas{i}}}
}
\bm{
= \sums{\kas{1}}\cdots\sums{\kas{\ac}}
\cas{1\kas{1}}\cdots\cas{\ac\kas\ac} \eu{\ia\brac{\eps{\ks1} + \cdots +
\eps{\kas{A}}}t/\hbar} \dket{\us{\kas{1}} \cdots \us{\kas{\ac}}}\;, 
}
where \brar{\us{\kas{1}} \cdots \us{\kas{\ac}}} is the basis of sp-states and 
\see{-0} can be rewritten as
\bee{det11}{
\dket{\psis\ac} = \sums{\ja} \calcsj \eu{\ia\ucs{\ja}\ta/\hbar} \dket{\calwsj}
}
where \ja\ is a characteristic integer for the \embm{nuclear energy}, corresponding to a
specific sequence of the enumerable set of components of {\psis\ac}. In this case we use
the notation (\ja) to designate the corresponding degenerate set of {nuclear
configurations} with energy \ucs{\ja}.

The grand canonical ensemble vector corresponding to the various nuclear states
\m{\dket{\psis\ac}} can also be written\cite{fetter} directly in terms of the
\embm{occupation numbers of all sp-states} defined in \ree{det3} as
\bee{det12}{ \dket{\psi}=\sums{\nas{1},\cdots,\nas{\infty}}
\dket{\nas{1}\cdots\nas{\infty}}\,,}
and if one defines the symbol ``\brac{\ks{1}\cdots\ks{\ac}}" for the set of all
configurations containing \ac\ and only \ac\ sp-states with non null occupations as
\bee{det13}{ \brac{\ks{1}\cdots\ks\ac} = \brar{
\brar{\ks1,\cdots,\ks\ac},\suchthat\ns{\ks1}=\cdots=\ns{\ks\ac}=1},
\mbox{\hspace{6pt} for given \m{\ac},}
}
where \m{\ks1,\cdots,\ks\ac} are also supposed to be ordered by increasing values, then
\see{-1} can be written as
\bee{det14}{ \dket{\psi} =
\sums{\cdots,\ks{1},\cdots,\ks\ac,\cdots}\dket{\ns{1},\cdots,\ns{\ka{1}},\cdots,\ns{\ka\ac},\cdots,\ns{\infty}}=
\sums\ac\sums{\ks{1},\cdots,\ks\ac}\dket{\brac{\ks{1}\cdots\ks\ac}}=
\sums\ac \dket{\psis\ac}
}

The Fock space operator of an excited nuclear sytem that describes the
corresponding grand canonical ensemble of nuclear states in terms of \embm{excitons},
with {\brac\pa} ``particles"  and {\brac\ha} ``holes" and arbitrary \ac,  and has
expected values over the states \ket{\psi} given by \see{-11} or \see{-10} for each
exciton type, can be defined, considering ``\pa" and ``\ha" as independent fermion
fields, by the expression\cite{obp}
\bee{det15}{ \fcs0 =\prods{\mu\nu}\brac{\asnu\auspnu+\xspnu\auspnu\asnu}
                                  \brac{\bsmu\buspmu+\xshmu\buspmu\bsmu}
 =\prods{\mu\nu} \fcs{p\nu}\fcs{h\mu}\;.  }

If one considers initially the simpler one fermion expression
\bee{det17}{ \fcs{0} =\prods{\la}\brac{\asl\auspl+\xspl\auspl\asl}
,}
it can be rewritten as
\beeplusthree{det18}{ \fcs0 = \prodsu{\la=0}{\infty}\brak{\brac{\asl\auspl}\cdots} +
\sigsu{\sa=1}{\infty}\brak{\prods{ (\ls1,\ls2)\choose \dif \sa}
\brac{\as{l1}\aus\dagger{l1}}\cdots\brac{\aus\dagger\sa\as\sa}\cdots\brac{\as{l2}\aus\dagger{l2}}
\cdots}\xs{ps}
}
{+ \sums{\sa1<\sa2}\brak{\prods{ (\ls1,\ls2,\ls3)\choose \dif (\sa1,\sa2)}
\brac{\as{l1}\aus\dagger{l1}}\cdots\brac{\aus\dagger{\sa1}\as{\sa1}}\cdots
\brac{\as{l2}\aus\dagger{l2}}\cdots\brac{\aus\dagger{\sa2}\as{\sa2}}\cdots
\brac{\as{l3}\aus\dagger{l3}}
\cdots}\xs{p\sa1}\xs{p\sa2}
}
{+\sums{\sa1<\sa2<\sa3}\brac{\cdots }\xs{p\sa1}\xs{p\sa2}\xs{p\sa3}
+\cdots }
{ = \sumsu{\nc=0}{\infty} \sums{\ss{\nc}}\brac{\prodsu{\ja=1}{\nc}\xs{p\ss{j}} }
\Pi{\brac{\ss\nc}}
}
where, for fixed \nc\ and configuration (\ss\nc), the Fock operator {\Pi\brac{\ss\nc}} is
given by
\bm{ \Pi{\brac{\ss\nc}}=
\prods{(\ls1,\cdots,\ls{\nc},\ls{\nc+1}) \choose \dif(\sa1,\cdots,\sa\nc)}
\brac{\as{l1}\aus\dagger{l1}}\cdots\brac{\aus\dagger{\sa1}\as{\sa1}}\cdots
\brac{\as{l\nc}\aus\dagger{l\nc}}\cdots\brac{\aus\dagger{\sa\nc}\as{\sa\nc}}\cdots
\brac{\as{l(\nc+1)}\aus\dagger{l(\nc+1)}}
\cdots }
with {\nc} terms of the type
{\brac{\aus\dagger{\sa\ja}\as{\sa\ja}}} among an infinite number of terms of the type
{\brac{\as{\la\ja}\aus\dagger{\la\ja}}}.

Then, it is clear that
\bee{det201}{ \Pi{\brac{\ss\nc}}\dket{\brac{\ks1\cdots\ks\ac}} =
\des{\nc,\ac}\d\rbk{\ss1\cdots\ss\nc}{\ks1\cdots\ks\ac}\dket{\brac{\ks1\cdots\ks\ac}}
}
and therefore
\beeplusfour{det19}{\fcs0\dket{\psi} =
\fcs0\sums\ac\dket{\psis\ac} }
{= \sums\nc\sums{\ss\nc}\brac{\prodsu{\ja=1}{\nc}\xs{p\ss{j}}}\Pi{\brac{\ss\nc}}
\sums\ac\sums{\ks\ac}\dket{\brac{\ks1\cdots\ks\ac}}
}
{= \sums{\nc,\ac}\sumst{\ss{\nc}}{\ks{\ac}}
\brac{\prodsu{\ja=1}{\nc}\xs{p\ss{j}}}
\des{\nc,\ac}\d\rbk{\ss1\cdots\ss\nc}{\ks1\cdots\ks\ac}\dket{\brac{\ks1\cdots\ks\ac}} = }
{= \sums{\nc,\ac}\sumst{\ss{\nc}}{\ks{\ac}}
\brac{\prodsu{\ja=1}{\nc}\xs{p\ss{j}}}
\des{\nc,\ac}\d\rbk{\ss\nc}{\ks\ac}\dket{\ks\ac} =
}
{= \sums\nc\sums{\ss1\cdots\ss\nc}\brac{\prodsu{\ja=1}{\nc}\xs{p\ss{j}}}\dket{\brac{\ss1\cdots\ss\nc}}
\eq\sums\nc\dket{ \widetilde{\psis\nc} }
}
which, by comparison with \see{-12} shows that \m{\fcs0} projects the grand canonical ensemble vector
\m{\dket{\psi}} of Eq.\ree{det12} into another linear combination of its various components, where
the coefficients have changed from {1} to
\bm{\brac{\prodsu{\ja=1}{\nc}\xs{p\ss{j}}},}
which are proportional (for given \nc) to the canonical ensemble \embm{probabilities} to
find the nuclear system in each of the component configurations
\dket{{\ss1\cdots\ss\nc}}.\cite{reif}

Eq.\see{-1} also implies that for two given independent grand canonical vectors
\bm{\sums{1}\dket{1}=\sums{\ncs{1}}\sums{\ks{\ncs{1}}}
\dket{\brac{\ks{1}\cdots\ks{\ncs{1}}}} \;\;\mbox{and}\;\;
\sums{2}\dket{2}=\sums{\ncs{2}}\sums{\ls{\ncs{2}}} \dket{\brac{\ls{1}\cdots\ls{\ncs{2}}}} }
results,
\beeplusthree{det221}{ \sums{1,2} \dbok{1}{\fcs0}{2} = \sums{1} \dbra1
\sums{\nc,\ncs{2}}\sums{\ss\nc,\ls{\ncs{2}}}
\brac{\prodsu{\ja=1}{\nc}\xs{p\ss{j}}}
\des{\nc,\ncs{2}} \d\rbk{\ss\nc}{\ls{\ncs{2}}}
\dket{\ls{\ncs{2}}} }
{ = \sums{\ncs{1},\nc,\ncs{2}}\sums{\ks{\ncs{1}},\ss{\nc},\ls{\ncs{2}}}
\brac{\prodsu{\ja=1}{\nc}\xs{p\ss{j}}}
\des{\nc,\ncs{2}}
\d\rbk{\ss{\nc}}{\ls{\ncs{2}}}\braket{\kas{\ncs{1}}}{\ls{\ncs{2}}}
}
{ = \sums{\ncs{1},\nc,\ncs{2}}\sums{\ks{\ncs{1}},\ss{\nc},\ls{\ncs{2}}}
\brac{\prodsu{\ja=1}{\nc}\xs{p\ss{j}}}
\des{\nc,\ncs{2}} \des{\ncs{1},\ncs{2}}
\d\rbk{\ss{\nc}}{\ls{\ncs{2}}}
\d\rbk{\ks{\ncs{1}}}{\ls{\ncs{2}}} }
{ = \sums{\ncs{1},\nc,\ncs{2}} \sums{\ks{\ncs{1}},\ss{\nc},\ls{\ncs{2}}}
\brac{\prodsu{\ja=1}{\nc}\xs{p\ss{j}}}
\braket{\ks{\ncs{1}}}{\ss{\nc}} \braket{\ss{\nc}}{\ls{\ncs{2}}}\;,
 }
therefore, \fcs{0} is algebraically equivalent to the unitary projector weighted by the
canonical probabilities of the microstes
\bm{\fcs{0} \rightleftarrow \sums{\nc,\ss{\nc}}
\brac{\prodsu{\ja=1}{\nc}\xs{p\ss{j}}} \ketbra{\ss{\nc}}{\ss{\nc}}  \;. }

Let \oc\ be an operator on the Fock space of the sp-states, which can modify the
nuclear configuration.  Then one can write the corresponding \embm{transition strength},
defined as the square of the transition moment \dbok{1}{\oc}{2} summed over all possible
transitions (``trans."), as\cite{hilborn}
\bn{ \scs{\oc} =  \sums{trans.} \absu{\oc}{2}
= \sums{1} \dbok{1}{\oc\ocu{\dag}}{1}
= \sums{1,2} \abs{\dbok{1}{\oc}{2}}\up{2}
= \sums{1,2} \dbok{1}{\oc}{2} \dbok{2}{\ocu{\dag}}{1}
\; }

\hst due to possible variation in the number of excitons and nuclear excitation \scs{\oc}
must be redefined in the framework of the grand canonical ensemble using the grand
canonical distribution, which can be written schematically as
\beeplusone{otwo}{ \absu{\oc}{2} = \dbko{\fcs0\oc\fcps{0}\ocu\dag} =
\sums{1}\dbok{1}{\fcs0\oc\fcps0\ocu{\dag}}{1} = \sumss{12}{34}
\dbok{1}{\fcs0}{2}
\dbok{2}{\oc}{3}
\dbok{3}{\fcps0}{4}
\dbok{4}{\ocu\dag}{1}
 \; }
{ = \sums{\nc}\sums{\ss\nc} \brac{\prodsu{\ja}{\nc}\xs{p\ss{\ja}}}
\dbok{\ss\nc}{\oc}{\rs{\mc}} \sums{\mc}\sums{\rs\mc}
\brac{\prodsu{\ka}{\mc}\xps{p\rs{\ka}}} \dbok{\rs\mc}{\ocu\dag}{\ss\nc}
}

In the case of two Fermion fields (e. g., ``particles" and ``holes") \fcs0 is given by
Eq.\see{-10}
\bee{det25}{ \fcs0 = \fcs{p}\fcs{h} =
\prods{\mu}\brac{\as\mu\aus\dag{\mu}+\xs{p\mu}\aus\dag{\mu}\as\mu}
\prods{\nu}\brac{\bs\mu\bus\dag{\mu}+\xs{h\mu}\bus\dag{\mu}\bs\mu} ,}
which can be rewritten, as we saw in \see{-11}, as
\bm{ \fcs\pa = \sigsu{\pa=0}{\infty}
\sums{\ss\pa}\brac{\prodsu{\ka=1}{\pa}\xs{\pa\ss\ka} } \Pi{\brac{\ss\pa}} }
and
\bm{ \fcs\ha = \sigsu{\ha=0}{\infty}
\sums{\rs\ha}\brac{\prodsu{\ja=1}{\ha}\xs{\ha\rs\ja} } \Pi{\brac{\rs\ha}} }
and, due to the properties of the single particle fermion operators, see \see{-8} and
\see{-4}, one can
also identify \m{\Pi\brac{\ss\pa}} and \m{\Pi\brac{\rs\ha}} with the components of the
respective projection operator,
\bee{projp}{ \Pi\brac{\ss\pa} = \dkb{\ss\pa}{\ss\pa}
 \mbm{ \ \ \ \ \ and \ \ \ \ \ \ \ }   \Pi\brac{\rs\ha} = \dkb{\rs\ha}{\rs\ha} }
and
\bee{projph}{
\Pi\brac{\ss\pa}\Pi\brac{\rs\ha}=\dkb{\ss\pa\rs\ha}{\ss\pa\rs\ha} .}
Then the analogous of \see{-7} becomes,
\bm{ \sums{1,2} \dbok{1}{\fcsp\fcsh}{2} = \sums{1,2} \dbra1
\sums{\pa\ha}\sums{\ss\pa\rs\ha}
\brac{\prodsu{\ka,\ja=1}{\pa\ha}\xs{\pa\ss{\ka}}\xs{\ha\rs{\ja}}
} \d\rbk{\ss\pa\rs\ha}{2}\dket{\brac{\ss\pa\rs\ha}} }
\bee{det26}{ = \sums{1,2} \sums{\pa\ha}\sums{\ss\pa\rs\ha}
\brac{\prodsu{\ka,\ja=1}{\pa\ha}\xs{\pa\ss{\ka}}\xs{\ha\rs{\ja}}}
\d\rbk{\ss\pa\rs\ha}{1}\d\rbk{\ss\pa\rs\ha}{2}
}
and
\bm{ \sums{1,2} \dbok{1}{\fcpsp\fcpsh}{2} =
\sums{1,2} \dbra1 \sums{\pp\hp}\sums{\ss\pp\rs\hp}
\brac{\prodsu{\la,\ia=1}{\pp\hp}\xps{\pa\ss{\la}}\xps{\ha\rs{\ia}} }
\d\rbk{\ss\pp\rs\hp}{2}\dket{\brac{\ss\pp\rs\hp}} }
\bee{det261}{ = \sums{1,2} \sums{\pp\hp}\sums{\ss\pp\rs\hp}
\brac{\prodsu{\la,\ia=1}{\pp\hp}\xps{\pa\ss{\la}}\xps{\ha\rs{\ia}}}
\d\rbk{\ss\pp\rs\hp}{1}\d\rbk{\ss\pp\rs\hp}{2}
}
and the analogous of \see{-10} and \see{-9} are
\bm{ \dbko{\fcsp\fcsh\oc\fcpsp\fcpsh\ocu\dag} =
\sums{1}\dbok{1}{\fcsp\fcsh\oc\fcpsp\fcpsh\ocu\dag}{1} }
\bee{det27}{ = \sumss{12}{34}
\dbok{1}{\fcsp\fcsh}{2}
\dbok{2}{\oc}{3}
\dbok{3}{\fcpsp\fcpsh}{4}
\dbok{4}{\ocu\dag}{1}
}
\bee{det28}{ = \sums{\pa\ha}\sums{\ss\pa\rs\ha}
\brac{\prodsu{\ka,\ja=1}{\pa\ha}\xs{\pa\ss{\ka}}\xs{\ha\rs{\ja}}}
\dbok{\ss\pa\rs\ha}{\oc}{\ss\pp\rs\hp}
\sums{\pp\hp}\sums{\ss\pp\rs\hp}
\brac{\prodsu{\la,\ia=1}{\pp\hp}\xps{\pa\ss{\la}}\xps{\ha\rs{\ia}}}
\dbok{\ss\pp\rs\hp}{\ocu\dag}{\ss\pa\rs\ha}
 }
which can be rewritten as
\bee{det29}{ \dbko{\fcsp\fcsh\oc\fcpsp\fcpsh\ocu\dag} =
\sumss{\pa\ha}{\pp\hp}\sumss{\ss\pa\rs\ha}{\ss\pp\rs\hp}
\brac{\prodsu{\ka,\ja=1}{\pa\ha}\prodsu{\la,\ia=1}{\pp\hp}
\xs{\pa\ss{\ka}}\xs{\ha\rs{\ja}}\xps{\pa\ss{\la}}\xps{\ha\rs{\ia}} }
\dbokt{\ss\pa\rs\ha}{\oc}{\ss\pp\rs\hp}{\ocu\dag} \;.
 }

Now we make the usual change of variables that defines the explicit connection with the
microscopic statistical parameters of the sp-states and also opens the possibility of the
Laplace transform interpretation
\bn{
 \xs{\pa\ss{\ka}}=\xa\eub{-\b\eps{\ss{\ka}}-\g\ms{\ss{\ka}}} }
\bn{
\xps{\pa\ss{\la}}=\xp\eub{-\bep\eps{\ss{\la}}-\gamp\ms{\ss{\la}}} }
\bn{
 \xs{\ha\rs{\ja}}=\ya\eub{-\b\eps{\rs{\ja}}-\g\ms{\rs{\ja}}}  }
\bee{det30}{
\xps{\ha\rs{\ia}}=\yp\eub{-\bep\eps{\rs{\ia}}-\gamp\ms{\rs{\ia}}} }
then \see{-1} can be rewritten as
\bn{ = \sumss{\pa\ha}{\pp\hp} \xu{\pa}\xpu{\pp}\yu{\ha}\ypu{\hp}
\sumss{\ss\pa\rs\ha}{\ss\pp\rs\hp}
\brac{\prodsu{\ka,\ja=1}{\pa\ha}\prodsu{\la,\ia=1}{\pp\hp}
\eub{-\b\eps{\ss{\ka}}-\g\ms{\ss{\ka}}}
\eub{-\b\eps{\rs{\ja}}-\g\ms{\rs{\ja}}}
\eub{-\bep\eps{\ss{\la}}-\gamp\ms{\ss{\la}}}
\eub{-\bep\eps{\rs{\ia}}-\gamp\ms{\rs{\ia}}} }
 }
\bee{det32}{ \dbokt{\ss\pa\rs\ha}{\oc}{\ss\pp\rs\hp}{\ocu\dag}
 }
or
\bee{det33}{ \dbko{\fcsp\fcsh\oc\fcpsp\fcpsh\ocu\dag} =
\sumss{\pa\ha}{\pp\hp} \xu{\pa}\xpu{\pp}\yu{\ha}\ypu{\hp}
\sumss{\ss\pa\rs\ha}{\ss\pp\rs\hp} \eu{-\b\uc-\g\mc-\bep\ucp-\gamp\mcp}
\dbokt{\ss\pa\rs\ha}{\oc}{\ss\pp\rs\hp}{\ocu\dag} }
where
\bee{det34}{ \uc=\sumsu{\ka,\ja=1}{\pa\ha}\eps{\ss{\ka}}+\eps{\rs{\ja}}
\m{\hspace{0.5cm}and\hspace{0.5cm}}
\mc=\sumsu{\ka,\ja=1}{\pa\ha}\ms{\ss{\ka}}+\ms{\rs{\ja}} }
\bee{det35}{
\ucp=\sumsu{\la,\ia=1}{\pp\hp}\eps{\ss{\la}}+\eps{\rs{\ia}}
\m{\hspace{0.5cm}and\hspace{0.5cm}}
\mcp=\sumsu{\la,\ia=1}{\pp\hp}\ms{\ss{\la}}+\ms{\rs{\ia}} }

As a shorthand practical notation that includes the essential features of the above
expressions one may define
\bm{ \dbko{\fcsp\fcsh\oc\fcpsp\fcpsh\ocu\dag} =  \rbk{\oc}{\ocu\dag}
= \sumst{12}{\uc\mc} \dbokt{\ss\pa\rs\ha}{\oc}{\ss\pp\rs\hp}{\ocu\dag} }
\bee{det33}{ = \sums{1}\sums{2} \eu{\brak{\uc\mc}}
\dbokt{\ss\pa\rs\ha}{\oc}{\ss\pp\rs\hp}{\ocu\dag} }
where
\bee{det341}{ \sums{1} =
\sumss{\pa\ha}{\pp\hp}\xu{\pa}\xpu{\pp}\yu{\ha}\ypu{\hp}
}
represents the sum over all possible numbers of excitons and
\bee{det341}{ \sums{2} = \sumss{\ss\pa\rs\ha}{\ss\pp\rs\hp}  }
represents the sum over all configurations for given exciton number and
\bee{det35}{ \eu{\brak{\uc\mc}} = \eu{-\b\uc-\g\mc-\bep\ucp-\gamp\mcp}\;,}
which is the non normalized gand canonical distribution function.

At last one may just drop the \sa's and \ra's and write
\bm{ \dbko{\fcsp\fcsh\oc\fcpsp\fcpsh\ocu\dag} = \rbk{\oc}{\ocu\dag} =
\sums{12}\eu{\brak{\uc\mc}}
\dbokt{\pa\ha}{\oc}{\pp\hp}{\ocu\dag},
}
which now has a precise meaning, where \ket{\pa,\ha}, \ket{\pp,\hp} represent the \emm{possible
nuclear configurations for given exciton numbers} ``{\pa,\ha}" and ``{\pp,\hp}".

In the case of the simple expected values (or ``first order momentum") of {\oc} the
expressions are totally analogous,
\bm{ \dbko{\fcsp\fcsh\oc} =  \dbko{\oc}
= \sumst{12}{\uc\mc}\dbok{\ss\pa\rs\ha}{\oc}{\ss\pa\rs\ha}
 = \sums{12} \eu{\brak{\uc\mc}}
\dbok{\ss\pa\rs\ha}{\oc}{\ss\pa\rs\ha} }
with the two sums given by
\bee{det36}{ \sums{1} =
\sums{\pa\ha}\xu{\pa}\yu{\ha}
\m{\hspace{0.5cm}and\hspace{0.5cm}}
\sums{2} = \sums{\ss\pa\rs\ha}  }
and
\bee{det37}{ \eu{\brak{\uc\mc}} = \eu{-\b\uc-\g\mc} }

Again one may drop the \sa's and \ra's to obtain the simplified expression
\bm{\brac{\oc} = \sums{12} \eu{\brak{\uc\mc}} \dbok{\pa\ha}{\oc}{\pa\ha}
}

Now, it is clear that the application of \emm{CAP} to the set of nuclear levels (
\uc\approx \embm{continuous}) on Eq.~\see{-7} will correspond to a very large
\embm{number of levels} per unit energy on \sigsil{2} and permit the approximate
replacement of the sum by an integral. The details of this procedure, its interpretation
and some consequences are analyzed next.

\subsecto{The connection with the Laplace transform }

The nuclear excitation, {\uc}, is a parameter that varies in a region defined by two
finite extremes
\bm{ \ucs{\min} \le \uc \le \ucs{\max}  }
where it takes a sequence of discrete values with degeneracy \cald(\ac,\uc,\mc), as
defined in the last section, with \ac\ interpreted as the total number of excitons,
{ \ac=\na=\pa+\ha.}
Similarly the total nuclear angular momentum projection also varies in a stepwise manner
between \m{\mcs{min}} and \m{\mcs{max}} as does any \embm{additive quantum number} of the
nuclear system. The following analysis is valid for these parameters, which correspond to
the extensive thermodynamic properties of the system.

For given numbers (\pa,\ha) there is in general a subset of the set \brac{\ss\pa\rs\ha},
of all nuclear configurations associated with (\pa,\ha), for which the quantum numbers
(\uc,\mc) take the same values and, by definition, the number of elements of this subset
is equal to the degeneracy of the corresponding nuclear state,
\bee{det38}{ \brar{\bracs{\ss\pa}{\ia},\bracs{\rs\ha}{\ia}} =
\brar{\brars{\ss1,\cdots,\ss\pa,\rs1,\cdots,\rs\ha}{\ia}, \,
\ia\bel\brar{1,\cdots,\cald(\ac,\uc,\mc)} }
\;.}
Then, for example, one can rewrite Eq.\see{-5} as
\bn{ \dbko{\fcsp\fcsh\oc} = \sums{12}
\eu{\brak{\uc\mc}}\dbok{\ss\pa\rs\ha}{\oc}{\ss\pa\rs\ha} } \bm{ = \sums{1}\mbox{\ }
\sigsu{\mc=\mcs{min}}{\mcs{max}}\mbox{\ } \sigsu{\uc=\ucs{\min}}{\ucs{\max}}
\sigsu{\ia=1}{\cald(\ac,\uc,\mc)}\eu{\brak{\uc\mc}}
\dbok{\bracs{\ss\pa}{\ia}\bracs{\rs\ha}{\ia}}{\oc}{\bracs{\ss\pa}{\ia}\bracs{\rs\ha}{\ia}}
}

The cumulative number of nuclear states can be very high even for not very high excitations,
as indicated by phenomenological calculations of nuclear level densities.\cite{tng} Then,
if the density is also high, it is reasonable to use of \emm{CAP} to replace the sum over
\uc\ by an integral in \see{0} and one can explore this possibility using an  \embm{ad
hoc} definition of nuclear density, inspired by the analysis at the beginning of this
section.

Note that for each discrete \uc\ in the sum in the RHS of \see{0}\m{\,} there are
\cald(\ac,\uc,\mc) states with the same energy and angular momentum, for which
\eu{\brak{\uc\mc}} has the same value.
Then, if \de\uc=\brac{\uc-\ucs{prev}} is the variation of the nuclear excitation between
its present and ``previous value" in the sequence of the discrete set to which \uc\
belongs, the corresponding approximate nuclear density for each index \ia\ on the last
sum of the RHS of \see{0} will be
\bn{\ome\brac{\ac,\uc,\mc}\approx \cald(\ac,\uc,\mc)/\de\uc \approx constant,}
and
\bm{ \cald(\ac,\uc,\mc)=
\sumsu{\ia=1}{\cald(\ac,\uc,\mc)} (1)
\approx \ome\brac{\ac,\uc,\mc} \de\uc =
\intsu{\ucs{prev}}{\uc}\ome\brac{\ac,\uc,\mc} (1) \da\uc.}
Then, replacing ``(1)" by an arbitrary integrable function ``(...)" and summing over all \uc\ gives,
\bm{\sumsu{\uc=\ucs{\min}}{\ucs{\max}}\sumsu{\ia=1}{\cald(\ac,\uc,\mc)} (...)
\approx \intsu{\uc=\ucs{\min}}{\ucs{\max}}\ome\brac{\ac,\uc,\mc} (...) \da\uc.}
and the sum over configurations with given a  number of excitons in \see{-2} becomes
\bm{ \sigsu{\mc=\mcs{min}}{\mcs{max}} \mbox{\ }\sigsu{\uc=\ucs{\min}}{\ucs{\max}}
\sigsu{\ia=1}{\cald(\ac,\uc,\mc)} \eu{\brak{\uc\mc}}  \approx
 \intsu{\uc=\ucs{\min}}{\ucs{\max}}  \eu{-\b\uc}
\brac{\sumsu{\mc=\mcs{min}}{\mcs{max}} \ome\brac{\ac,\uc,\mc} \eu{-\g\mc}} \da\uc
}
where the definition \see{-7} for \m{\eu{\brak{\uc\mc}}} was used.

Now, for the nuclear excitation energies the minimum is the ground state,
corresponding to {\ucs{\min}=0}, and the maximum is unbounded and one can take,\cite{w71}
\bee{det41}{ \ucs{\max} \approx \infty \mbox{,\hspace{0.1cm} with good approximation,} }
{if} the contribution for energies above \ucs{\max} can be neglected. 
In this case, \see{-1} becomes the Laplace transform of the part of the integrand
inside the parenthesis.

More generally, one can rewrite \see{-4}, without \emm{CAP}, as
\bm{ \dbko{\fcsp\fcsh\oc}= \sums{1} \sumsu{\mc=\mcs{min}}{\mcs{max}} \sumsu{\uc=0}{\infty}
\sums{\al} \eu{\brak{\uc\mc}}
\da\sub{\a}(\pa,\ha,\uc,\mc)
\dbkos{\oc}{\al}(\pa,\ha,\uc,\mc)
}
where \a\ indicates all configurations for which
``\dbok{\pa\ha\uc\mc\!}{\oc}{\pa\ha\uc\mc}" has the same value, i. e., the configurations
\embm{degenerated} with respect to the action of \oc\, or the observation of the physical
quantity represented by \oc, therefore with the same expected value \dbkos{\oc}{\al},
and ``\da\sub{\a}(\pa,\ha,\uc,\mc)"
is the corresponding degeneracy of states, with sum \embm{exactly} equal to
\cald(\ac,\uc,\mc)
\bm{ \sums{\a} \da\sub{\a}(\ac,\uc,\mc) = \cald(\ac,\uc,\mc) \;.}
The important point of expression \ree{2.67} is that it can already be used for
numerical calculations, but the statistical models prefer to employ the approximated
Laplace transform formalism, \embm{as if it was equivalent} to the direct microscopic
description.  

Similarly to \ree{2.65}, Eq.\ree{2.67} can be approximated, using \emm{CAP}, as
\bm{ \dbko{\fcsp\fcsh\oc} \approx \sums{1} \sumsu{\mc=\mcs{min}}{\mcs{max}}
\intsu{0}{\infty} \da\uc \eu{-\b\uc-\g\mc} \sums{\al} \om\sub{\a}(\pa,\ha,\uc,\mc)
\dbkos{\oc}{\al}(\pa,\ha,\uc,\mc)
}

\vhst where the nuclear density is defined as
\bn{\oms{\al}\brac{\ac,\uc,\mc}\approx\da\sub{\a}(\pa,\ha,\uc,\mc)/\de\uc \approx constant,}
for each \al, and one can rewrite
\bm{ \da\sub{\a}(\pa,\ha,\uc,\mc) \approx \oms{\al}\brac{\ac,\uc,\mc} \times \brac{\uc-\ucs{prev}}
\approx \oms{\al}\brac{\pa,\ha,\uc,\mc} \de\uc \;,}
corresponding to the replacement of discrete degeneracy \das{\a} by continuous level
density \oms{\a} and, for given number of excitons \m{(\pa,\ha)}, one has
\bm{ \sumsu{\mc=\mcs{min}}{\mcs{max}}
\intsu{0}{\infty} \da\uc \eu{-\b\uc-\g\mc}
\sums{\al} \om\sub{\a}(\pa,\ha,\uc,\mc)
\dbkos{\oc}{\al}(\pa,\ha,\uc,\mc)\;.
}

\vhst Therefore, by definition,
\bm{ \dbko{\fcsp\fcsh\oc} \app \lapt{
\sumsu{\mc=\mcs{min}}{\mcs{max}} \eu{-\g\mc} \sums{\a}
\om\sub{\a}(\pa,\ha,\uc,\mc)
\dbok{\pa\ \ha\ \uc\mc\!}{\oc}{\pa\ \ha\ \uc\mc}\sub{\a}
} \;,}

\vhst where the symbol \lapt{X} indicates the Laplace transform of (\xc).

In particular, in the simplest case of the identity operator, \oc=\mbm{1}$\!$, all expected
values are equal ``1" and \dbko{\fcsp\fcsh} becomes essentially the Laplace transform of
the state density, then
\bm{ \sumsu{\mc=\mcs{min}}{\mcs{max}}\eu{-\g\mc} \om(\pa,\ha,\uc,\mc) \app
 \call\up{-1}\brar{\dbko{\fcsp\fcsh}} =
\call\up{-1}\brar{\prods{\mu}\brac{1+\xs{\pa\mu}}\prods{\nu}\brac{1\pl\xs{\ha\nu}}}
 \;. }
This result is \embm{equivalent} to the traditional one of \ree{2.8}, if one
includes the angular momentum projections in the definition of the integration variable
\ya\ as in \see{-28},
because the RHS of Eq.\ree{2.8} can be interpreted, using \emm{CAP}, as the \emm{inverse}
Laplace transform of the grand canonical generating function.\cite{w71,b68}


On the other hand, one must notice that, in general,
\see{-65} is \embm{not} equivalent to \call\up{-1}\brar{\dbko{\fcsp\fcsh}}, that is,
the traditional definition of \see{-65} and the use of the Laplace transform for the
description of the physical obervation of \oc\ is {approximate},  while Eq.\see{-11} is
exact, assuming the grand canonical description of nuclear states with microstates
constrained only by the IPM (Eq.\ree{IPM}).

In addition, we have shown above that the microscopic formalism directly yields the
desired expressions corresponding to \embm{the inverse} of the Laplace transform of the
expected values of the interacting operators, in terms of \dbkos{\oc}{\al} and
the corresponding density of ``final available states" (i. e., the physically allowed
final states associated with the ``observation" or ``measurement" of \oc), \embm{without
having to actually evaluate them}.

Therefore, the straightforward and natural results of the \DFM\ are obscured by the
approximate, non necessary, intrincate and microscopically imprecise traditional
formalism of the Laplace transform, which should be avoided in \pe\ calculations. 
More reasons to avoid the Laplace transform formalism are given in Ref.\cite{fbgarXiv1}.

\subsecto{Discrete functions of the configurations}



Regarding the idea of nuclear degeneracy, one notices that in \see{-11} all the expected values
\bn{ \dbok{\bracs{\ss\pa}{\ia}\bracs{\rs\ha}{\ia}}{\oc}{\bracs{\ss\pa}{\ia}\bracs{\rs\ha}{\ia}} \; }
correspond to configurations with energy \uc\ and total degeneracy \cald(\ac,\uc,\mc), which we will
designate by (\ua)

\bm{(\ua) = \brar{(\sas{\pa\ia},\ras{\ra\ia}); \ia=1,\cald(\ac,\uc,\mc)}
 \;, }
with \ac=\na=\pa+\ha\ and given \mc.

If the operator \oc\ describes, for example, the measurement of the \emm{intrísic spin}
of the nuclear state it will have in general a sequence of different \emm{discrete
values} for the different elements of (\ua)
and the expected values, \dbko{\oc}, will also be degenerated, i. e., in general there will be more than one
configuration for each value of the total spin. Then, one may write
\bm{ \sigss{(\ua)}{(conf)}\dbkos{\oc}{\ua}=
\sumsu{\ia=1}{\cald(\ac,\uc,\mc)}\dbok{\bracs{\ss\pa}{\ia}\bracs{\rs\ha}{\ia}}{\oc}{\bracs{\ss\pa}{\ia}\bracs{\rs\ha}{\ia}}=
\sumsu{\ia=1}{\cald(\ac,\uc,\mc)}\dbkos{\oc}{\ia}=
\sigssu{\be=1}{(spins)}{\bes{max}}\dbkos{\oc}{\be} \das{\be}(\ac,\uc,\mc)
}
where \brar{\be=1,\cdots,\bes{max}} is a sequence of \emm{integers} in biunivocal (one to
one) correspondence with the discrete set of ``spin values" \brars{\dbkos{\oc}{\be}}{\be}
and \das{\be}(\ac,\uc,\mc) is the ``spin degeneracy" satisfying,
\bm{ \sumsu{\be}{\bes{max}} \das{\be}(\ac,\uc,\mc) = \cald(\ac,\uc,\mc)}

This idea can be straightforwardly generalized to the ``measurement" of a quantity that changes the
total number of excitons, \ac=\na=(\pa+\ha), for example:
\oc=\sumsne{\al}\oc\sub{\al}\asu\a\dag\as\a,
where ``\al" indicates an exciton state of ``particle" type. Then, the
expected values in \see{-14} would select all configurations in (\ua), Eq.\see{-2}, that
have one sp-state ``\al" in them.

The number of configurations for which the
destruction of an sp-state ``\al" has non null expected value, define the degeneracy of
nuclear states containing ``\al". If this number is designated by \da\sub{\al}(\ac,\uc,\mc),
the resulting set of nuclear configurations after the ``observation" of \aas{\a}, the
\embm{intermediary state}, would have \emm{the same degeneracy}, but the corresponding
grid of nuclear energies and angular momenta would be displaced by \eps{\al} and
\mas{\al} respctively, therefore
\bm{ \da\sub{\al}(\ac-1,\uc-\eps{\al},\mc-\mas{\al}) = \da\sub{\al}(\ac,\uc,\mc) \;.}

In this context the introduction of \emm{CAP} naturally brings the idea of convolution
between states of different levels of ``complexity" (number of excitons) during the
\pe-stage as it transforms sums of discrete sets of expected values times the respective
degeneracies, Eq.\see{-2}, into integrals involving the corresponding nuclear densities.

Notice though that this connection with the idea of convolution is somewhat artificial,
as it is not related with any actual physical connection between configurations with
different number of excitons.

The result of Eq.\see{-0} is only an aspect of the combinatorial relation between the
different sets of
configurations, where a sp-state is ``selected" by the microscopic
interactions in the set of configurations with greater number of excitons, producing a
new set that
happens to have the \embm{same number} of components,
in which the selected sp-state is not present. 
Therefore, it is important to notice that the connection between the two sets of
configurations in Eq.\see{-0} is exclusively numerical, 
not physical. 

The sums and integrals have ranges defined by the \embm{available states},
i. e.  the states that give non zero expected value for the measured quantity. For
example, in \see{-7} the density \oms{\a}(\ac,\uc,\mc) corresponds to available nuclear
states associated with the operation or measurement of \oc, among the set of microstates with
given \ac, \uc\ and \mc.

On the other hand, the set of configurations for each sp-state ``\a" is not directly
connected with the nuclear excitation energy, but if the density of states is high all
configurations with energies between two given energies, \uc\ and \ucs{prev}, can be
considered as approximately having ``energy \uc" and the variation over \al\ can be associated with
densities corresponding to \uc.

In particular, when one sums over an sp-state ``\al" the previous discussion shows that
it could have some degeneracy and, therefore, it is not equivalent to the sum over
\eps{\al}. This idea is further developed in Ref.\cite{fbgarXiv1} where the sum over
\eps{\a} is reduced to a convolution of nuclear densities, i. e., the sp-energies
have an \embm{associated density} that, in the \emm{CAP} limit, naturally defines
a convolution with the nuclear density that appears in the expression of the expected
values.

Notice that the sum over the total nuclear spin projection \mc, and the corresponding
sums over single particle angular momenta, \mas{\a}, \embm{cannot} be transformed in the
same way as {\uc} and \eps{\al}, using \emm{CAP}, because {the interval between sucessive
values of {\mc} or \mas{\a} is never smaller than {0.5} and therefore it cannot be
considered as an infinitesimal even for a large number of configurations}.

In \ref{section.4} we apply of the above formalism for the calculation of the transition
strengths (TS) of the PE nuclear Hamiltonian and for estimation of \pa-induced reactions on
\fe{56}.

\secto{Phenomenological Statistical description}

A complementary description to the
one presented in the previous section is the phenomenological statistical description of
\pe-states\cite{k72,b68,w71,w69} and the \oxm,\cite{kon04,k8586} that forms the
theoretical basis of the nuclear data model code \tn.\cite{tng}

This model describes the particles in the excited nucleus as belonging to a degenerated
Fermi gas, moving semi-classically, nearly independently of each other, and belonging to
highly degenerated nuclear levels with small interspacing between them, under the action of
an average potential of residual interactions (``mean field"), similarly to a system of
rigid macroscopic objects that continuously collide until one or more are eventually
``emitted" from the system, corresponding to the nuclear decay.
Only \na, \pa\ and \al\ particle emissions are considered in the \tn\ code.

Any excited particle can escape at any time, but is also subjected to quantum
kinematic constraints defined by the barrier of the attractive mean field potential and the
exclusion principle for fermions that may preclude emission.
Then, the excited nuclear state can be characterized at any time by its quasi-equilibrium
set of configurations of excitons (\qe\ state), that is the state in which all
configurations at each stage of the increasing complexity chain, i. e. for given (\na,\uc),
have equal probability of occurrence.

The \qe\ hypothesis is criticized in the work of Pompeia and Carlson\cite{pc} and found
unecessary. They suggest the replacement of the \oxm\ by a more general ``natural model",
but the results of the \DFM\ presented in this paper could invalidate this conclusion to some
extent, as we will see in the next section.

When the excitation energy of the residual nuclide is too low to permit more
\pe\ emissions the pre-equilibrium calculations stop and the residual
nucleus is considered to be at the final ``equilibrium" state, or \embm{compound nucleus}
state (\cn).  The \cn\ is then described by the nuclear liquid drop model for the
remaining de-excitation process, called the "evaporation" stage (EVAP).\cite{bethe36,haufesh}

The temporal evolution of the \qe\ state as a function of the nuclear excitation energy,
\uc, from an initial state \ket{\nas{\ia},\uc} to a final \ket{\nas{\fa},\uc}, is
described by the total transition probability per unit time or transition rate (TR),
\lam(\nas{\ia},\nas{\fa},\uc).

If one assumes that the residual potential, \vc(\uc),
does not change importantly during the transition, that \ket{\nas{\ia},\uc} and
\ket{\nas{\fa},\uc} are {stationary} and described approximately by the eigen-vectors of
the unperturbed Hamiltonian, and that the density of final states is
sufficiently high, then \lam(\nas{\ia},\nas{\fa},\uc) can be estimated using Fermi's
golden rule,\cite{semiclasssem}
\bee{e1}{\label{e1}
 \lam(\nas{\ia},\nas{\fa},\uc) = \frac{2\pi}{\hbar} \int
\absu{\dbok{\nas{\fa}}{V(\ucp)}{\nas{\ia}}}{2}
   \delta(\ucp - \uc ) \om(\nas{\fa},\ucp) d\ucp ,
}
where \om(\nas{\fa},\ucp) is the density of final nuclear states with energy \ucp\ and
exciton number \nas{\fa}.
The single particle states and the residual potential are usually defined
self-consistently and the final states are assumed to be obtained from the initial ones
exclusively through two-particle transitions, i. e., \delc\na=0 or \delc\na=\pm2.
If \ket{\na,\uc,\ta} is a given configuration of sp-states (nuclear state) with \na\
excitons and excitation \uc\ at the time \ta, then it belongs to a \embm{class of nuclear
states} defined by \brar{\na,\uc}, usually called an ``exciton class" because \uc\ is
supposed to be constant between particle emissions, even if \na\ varies.

Then, denoting by \pc(\na,\uc,\ta) the probability to find the nuclear system in the
state \ket{\na,\uc,\ta}, the
time evolution of \pc(\na,\uc,\ta) is governed by a master equation given
explicitly in terms of the different exciton transitions and possible particle emissions as,
\bm{ \frac{\da\pc(\na,\uc,\ta)}{\da\ta} =
\sums{\nu}\frac{\da\pcs{\nu}(\na,\uc,\ta)}{\da\ta} \;. }
where \nu\ indicates the different types of emitted particle and an independent master
equation is defined for the probability associated with each possible emission, in
agreement with the basic assumptions of the original \EXM\ theory of
Miller\embm{et.al.}\cite{g66,pc},

\vspace{0.25cm}
\bn{ \label{eqstate}  \frac{\da\pcs{\nu}(\na,\uc,\ta)}{\da\ta} =
\brak{\pcs{\nu}(n-2,U,t) \brac{\frac{p-1}{p}} + \frac{\fas{\nu}(p)}{p} \pc(n-2,U,t) }
  \lambda_{+}(n-2,\na,U) \; }
\bn{ + \lams{0}(\na,\na,\uc,\ta) \pcs{\nu}(\na,\na,\uc,\ta) +
\lams{-}(\na+2,\na,\uc,\ta) \pcs{\nu}(\na+2,\na,\uc,\ta) \; }
\beq \mbox{\hsp{30pt}}
 - \pcs{\nu}(n,U,t) \brak{ \lamc(n,U) + \intsu{\vcsu{\nu}{c}}{U-B_{\nu}}
\lamsu{\nu}{c}(\na,\uc,\eps{\nu}) \da\eps{\nu} }  \;,
\lee{eqstate}
where \pa\ is the number of particles, \lamc(\na,\uc) is the total transition rate for
nuclear states belonging to a given exciton class \brar{\na,\uc}, summed over
all possible inter-class transitions, from \ket{\nap,\uc} to \ket{\na,\uc},
\beq
\lamc(\na,\uc) = \sums{\nap} \lams{\delc\na}(\na,\nap,\uc)\;,
\lee{ceminc6}
\bcs{\nu} is the binding energy of the emitted particle in the compound nucleus and
\vcus{c}{\nu} is the Coulomb barrier before the emission.
In \tn\ the terms in \lams{0} are neglected in \ree{3.3} and \ree{3.4}, corresponding to
the hypothesis of attainment of ``perfect equilibrium" for each exciton class of the \pe\
process, or the IPM description in each class.
This corresponds to the assumption that, for given \brar{\na,\uc},
single particle transitions \embm{do not} happen and all nuclear configurations have
the same probability, which are usual \EXM\ hypotheses.

In \ree{eqstate}, the transition rates are usually given by an approximated form of \ree{e1}
\beq
\lambs{\delc\na}(\nas{\ia}+\delc\na,\nas{\ia},\uc) = \frac{2\pi}{\hbar}
\absu{\mc}{2}  \oms{\delc\na}(\nas{\ia}+\delc\na,\uc)
\lee{ceminc7}
where \nas{\ia} is the initial exciton number, \delc\na=0,\pm2, and
\abs{\mc} is a phenomenological average absolute value for the interacting potential, or
nuclear mean-field,
supposed independent of \uc\ and the type of transition.
The explicit expression for the state density is given by \cite{pc,wplb70}
\bm{ \om\of{\pa,\ha} = \frac{\ga[\ga\uc-\cc(\pa,\ha)-\delcs{pair}]\up{\pa+\ha-1}}{\pa!\ha!(\pa+\ha-1)!}
\fa\of{\na,\uc} \; }
where \cc(p,h) is the approximate correction factor associated with
Pauli's exclusion principle,\cite{w71,pc}
\bee{ccph}{ \cc(p,h)=\frac{1}{4}\brac{p^2 + h^2 - p - h} \,\, ,}
\delcs{pair} is an additional term to account for pairing among sp-states\cite{fu84} and
\fa\of{\na,\uc} is a term to account for the limited depth of the nuclear potential
well.

The density of available states for transitions with \delc\na=0 is given
by,\cite{w71,fu1}
\bm{ \oms{0}(\pa,\ha,\uc) \approx (\frac{\ga}{2}) \frac{(\ga\uc-\cc(p,h))}{(\pa+\ha)}
 (\pa(\pa-1)+\ha(\ha-1)+4\pa\ha)
\;, }
and for transitions that decrease the number of excitons by 2, by
\bm{ \oms{-}(\pa,\ha,\uc) \approx (\frac{\ga}{2}) \pa\ha(\pa+\ha-2)
 \;. }
where \ga\ is the density of sp-levels in the uniform-spacing model.

The explicit expression for the density of final states associated with \delc\na={+2} can
be written in general terms as\cite{pc,wplb70}
\bn{\om(\pa,\ha,\uc)\oms{+}(\pa,\ha,\uc)= \om(\pa+1,\ha+1,\uc)\oms{-}(\pa+1,\ha+1,\uc)
\; }
\bm{\approx \frac{(\pa+1)(\ha+1)(\pa+\ha)}{2} \ga \om(\pa+1,\ha+1,\uc)
\;, }
Then,
\bn{ \frac{\om\of{\pa+1,\ha+1}}{\om\of{\pa,\ha}} =
\frac{\fa\of{\ha+1,\uc}}{\fa\of{\ha,\uc}}
\frac{\ga[\ga\uc-\cc(\pa+1,\ha+1)-\delcs{pair}]\up{\pa+\ha+2-1}/(\pa+1)!(\ha+1)!(\pa+\ha+2-1)!}
{\ga[\ga\uc-\cc(\pa,\ha)-\delcs{pair}]\up{\pa+\ha+1}/\pa!\ha!(\pa+\ha-1)!}
\; }
\bn{ \approx  [\ga\uc-\cc(\pa,\ha)-\delcs{pair}]\up{2}
\frac{\pa!\ha!(\pa+\ha-1)!}{(\pa+1)!(\ha+1)!(\pa+\ha+2-1)!}
\; }
\bn{ = \frac{[\ga\uc-\cc(\pa,\ha)-\delcs{pair}]\up{2}}{(\pa+1)(\ha+1)(\pa+\ha)(\pa+\ha+1)}
\;, }
where we have used

\bn{ \frac{\ga[\ga\uc-\cc(\pa+1,\ha+1)-\delcs{pair}]\up{\pa+\ha-1}}
{\ga[\ga\uc-\cc(\pa,\ha)-\delcs{pair}]\up{\pa+\ha-1}} \approx 1
\; }

\hst and \fa(\pa,\ha,\uc)\approx\fa(\pa+1,\ha+1,\uc).
Then, comparing with \see{-5} results,
\bn{ \brac{\frac{2\pi\mcu{2}}{\hbar}}  \omu{+}(\pa,\ha,\uc) =
\brac{\frac{2\pi\mcu{2}}{\hbar}} \brac{\frac{\ga}{2}}
(\pa+1)(\ha+1)(\pa+\ha)\frac{\om\of{\pa+1,\ha+1}}{\om\of{\pa,\ha}} =
\; }
\bm{   \approx \brac{\frac{2\pi\mcu{2}}{\hbar}} \brac{\frac{\ga}{2}}
\frac{[\ga\uc-\cc(\pa,\ha)-\delcs{pair}]\up{2}}{(\pa+\ha+1)} =
\lams{+}(\pa,\ha,\uc)
\;. }

The rate of emission of \m\nu-particles per unit time and energy in Eq.\ree{3.3},
\lamsu{\nu}{c}(\na,\uc,\eps{\nu}),  is estimated independently with the help of the
principle of detailed balance\cite{k71}

\beq \label{ceminc8}
\lamsu{\nu}{c}(\na,\uc,\eps{\nu}) =
\brac{ \frac{2 s_{\nu} +1}{\pi^2 \hbar^3} } \mu_{\nu} {\cal A}_{\nu}(n)
\sigma_{\nu}(\epsilon_{\nu}) \epsilon_{\nu} \frac{\omega_{\nu}
(p-1,h,U-B_{\nu}-\epsilon_{\nu})}{\omega(p,h,U)} ,\lee{ceminc8}

\hst where \eps{\nu} is the kinetic energy of the emitted particle, \m{s_{\nu}} is its
spin and \m{\mu_{\nu}} is its reduced mass relative to the rest of the nuclear system.
The semi-phenomenological factor \cala\sub{\n}(\na) ensures that the emitted particle is
of type \nu\ and defined consistently between the \pe\ and the \cns stages, in a
procedure equivalent to the parameterization of Kalbach.\cite{kalbach210}

\secto{Comparison of the \DFM\ with the \EXM}

The transition strengths of the \DMF, described in \ref{section.2} and
Refs.\cite{fbgarXiv1,obp}, are calculated by code \trans (TST) 
for each possible microscopic process between excitons and summed over degenerate states,
for each \embm{class} of nuclear states defined by \na.

The model space is defined by the Harmonic Oscillator (H.O.) basis of single
particle states directly in terms of the hypergeometric function\cite{brussbook}, with
the usual addition of a strong spin-orbit term  to modulate the fixed inter-spacing of
the kinetic energy levels of the H.O. to try to adjust the resulting structure of nuclear
levels to the observed one.\cite{ring_shuck}

Another useful option is to define \embm{also}  the modulation of levels
semi-phenomenologically, by considering, for example, an arbitrary \embm{fixed
inter-spacing} between nuclear levels while keeping the other quantum numbers identical
to the H.O. basis, but the modulation could in principle be arbitrary. This assumption
corresponds to the approximation of considering the structure of the complex excited
nuclear states as \embm{unknown}, but similar enough to the ground state to be well
described by one additional phenomenological parameter.

Note that both definitions are
 {not} obtained from the direct solution of the nuclear many-body problem and, therefore,
 they represent an intrinsic {deficiency}, or inconsistency, of the present definition
 of the \DFM.
This inconsistency could be minored by adopting a more realistic
description for the nuclear potential, like the Woods-Saxon one,\cite{wsax54} or by
attempting a direct definition of the nuclear mean field from an approximated microscopic
solution of the many-body nuclear problem, like the Hartree-Fock approach.\cite{fetter}

The Fermi energy, \eps{\fc}, is defined as the energy of the last occupied single
particle state in the nuclear ground state and the energies of ``particles", are defined
as the negative of their binding energy in the nuclear system,
\bm{ \eps{\pa} = \eps{\fc} - \eps{\la} \;, }
where \eps{\la} is the energy of the sp-state belonging to sp-level ``\la", with maximum
binding, \eps{\pa,\max}, given by the complete occupation of all ``\ac" lower sp-states,
where \ac\ is the mass number. The energy of the ``holes" is
\bm{ \eps{\ha} = \eps{\la} - \eps{\fc} \; }
and all unoccupied sp-states above \eps{\fc} are considered possible hole states.

We neglect the hole states belonging to the level with energy \eps{\fc}.

We have made no restrictions regarding which ``particle" states are ``excitable" and
considered all different possible configurations of \pa-particles and \ha-holes that can
be distributed over the corresponding sp-states of the model spaces, as defined by the
above rules.

\subsecto{The problem of large degeneracies}

Following previous works \cite{pc,fbgarXiv1,wplb70} we have adopted the
``never-come-back" approximation, i. e., restricted the possible transitions to
\delc\na=0,+2, and considered the cases, listed in
\iftth \ret{Table_I}\else\ret{table.1}\fi,
for microscopic transitions in which the number of excitons increases by 2
and the transitions, listed in
\iftth \ret{Table_II}\else\ret{table.2}\fi, that keep the number of excitons constant.
The ``representations" of each transition follow the original definition of
Ref.\cite{obp} and the possible transitions are similar to those proposed in
Ref.\cite{wplb70}.

\vsp{2.3cm}
\tagn{table.1}{Table I}
\tagn{tablen.1}{I}
\tagn{Table I}{Table I}
\vspace{-3.50cm}
\begin{center} 
\begin{tabular}{p{2.2in}p{2.2in}p{2.2in}}
\multicolumn{3}{c}{\;\hspace{0.50in}\;\bf Table I - Microscopic transitions with \delc\na=+2.\;\hspace{0.50in}\;}\\\cline{1-3}
\end{tabular}  
\iftth  
\begin{tabular}{p{2.2in}p{2.2in}p{2.2in}}
\multicolumn{3}{c}{---------------------------------------------------------------------------------------------------}\\
\end{tabular}  
\fi  
{\sz 
\iftth  
\begin{tabular}{@{\qq}p{0.7in}p{1.0in}p{4.9in}}
{\bf case} & {\bf representation}  &  \hspace{90pt}{\bf description} \\
\else  
\begin{tabular}{@{\qq}p{0.5in}p{1.0in}p{3.1in}}
\hspace{10pt}{\bf case} & {\bf representation}  &  \hspace{70pt}{\bf description} \\
\fi  
\qq 1    &  (2100\leftarrow 1000)   &   proton-h-propagates  and   proton-ph-pair is created \\
\qq 2    &  (1011\leftarrow 1000)   &   proton-h-propagates  and  neutron-ph-pair is created \\
\qq 3    &  (1110\leftarrow 0010)   &  neutron-h-propagates  and   proton-ph-pair is created \\
\qq 4    &  (0021\leftarrow 0010)   &  neutron-h-propagates  and   proton-ph-pair is created \\
\qq 5    &  (1200\leftarrow 0100)   &   proton-p-propagates  and   proton-ph-pair is created \\
\qq 6    &  (0111\leftarrow 0100)   &   proton-p-propagates  and  neutron-ph-pair is created \\
\qq 7    &  (1101\leftarrow 0001)   &  neutron-p-propagates  and   proton-ph-pair is created \\
\qq 8    &  (0012\leftarrow 0001)   &  neutron-p-propagates  and  neutron-ph-pair is created \\
\end{tabular}
}
\iftth  
\begin{tabular}{p{2.2in}p{2.2in}p{2.2in}}
\multicolumn{3}{c}{---------------------------------------------------------------------------------------------------}\\
\end{tabular}  
\fi  
\end{center}
\vspace{1.50cm}

For given exciton class \brar{\na,\uc}, the nuclear configurations can be described by
a sequence of \na\ indices that characterizes their corresponding set of exciton
sp-states. The latter are given in this work by the full description of the H.O. basis,
\ket{\e,\sa,\mas{\sa},\la,\mas{\la},\t}, where \t\ is the isospin, \e\ the sp-energy,
(\sa,\mas{\sa}) the spin and spin projection and (\la,\mas{\la}) the orbital angular
momentum, etc..

Independently of the specific values of the quantum numbers characterizing sp-states,
they can always be in \embm{biunivocal} (one to one) relationship with the
``particle" and ``hole" states of the model space, therefore defining a sequence of
unique indices to describe these states,
\{\nas{\pa\ha},\nas{\na\ha},\nas{\pa\pa},\nas{\na\pa}\},
as follows: if the total number of \embm{excitable} proton-hole states (\pa\ha) in the
model space is \mas{ph}, the total number of \embm{excitable} neutron-hole states
(\na\ha) is \mas{nh}, neutron-particles (\na\pa) \mas{np} and
proton-particles (\pa\pa) \mas{pp}, then one can define
these indices as
\bm{ 1\le\nas{\pa\ha}\le\mas{ph}\;, }
\bm{ \mas{ph}\le\nas{\na\ha}\le\mas{ph} + \mas{nh}
\;, }
\bm{ \mas{ph} + \mas{nh}\le\nas{\pa\pa}\le\mas{ph} + \mas{nh} + \mas{pp},
\;, }
and
\bm{ \mas{ph} + \mas{nh} + \mas{pp}\le\nas{\na\pa}\le
\mas{ph} + \mas{nh} + \mas{pp} + \mas{np}
\;, }
and an arbitrary configuration with, for example, 3 excitons consisting of one proton-hole
and a neutron particle-hole pair can be uniquely indicated by the sequence
(\nas{\pa\ha},\nas{\na\ha},\nas{\na\ha});
a configuration with 4 excitons, being one particle-hole proton pair and
one particle-hole neutron pair, can be indicated by
(\nas{\pa\ha},\nas{\pa\pa},\nas{\na\ha},\nas{\na\ha}), etc..


\vsp{7.3cm}
\vspace{-1.00cm}
\tagn{table.2}{Table II}
\tagn{tablen.2}{II}
\tagn{Table II}{Table II}
\vspace{-7.00cm}
\begin{center} 
\begin{tabular}{p{2.2in}p{2.2in}p{2.2in}}
\multicolumn{3}{c}{\;\hspace{0.45in}\;\bf Table II - Microscopic transitions with \delc\na=0.\;\hspace{0.80in}\;}\\\cline{1-3}
\end{tabular}  
{\sz 
\iftth  
\begin{tabular}{@{\qq}p{0.7in}p{1.0in}p{4.9in}}
{\bf case} & {\bf representation}  &  \hspace{90pt}{\bf description} \\
\else  
\begin{tabular}{@{\qq}p{0.5in}p{1.0in}p{3.1in}}
\hspace{12pt}{\bf case} & {\bf representation}  &  \hspace{70pt}{\bf description} \\
\fi  
\qq  9  &   (1100\leftarrow 0011)  & neutron-ph-pair is destroyed and proton-ph-pair is created \\
\qq 10  &   (2000\leftarrow 2000)  &      scattering of two proton-h excitons \\ 
\qq 11  &   (1010\leftarrow 1010)  &      scattering of a neutron-h and a proton-h \\ 
\qq 12  &   (0020\leftarrow 0020)  &      scattering of two neutron-h \\ 
\qq 13  &   (0200\leftarrow 0200)  &      scattering of two  proton-p \\ 
\qq 14  &   (0101\leftarrow 0101)  &      scattering of a neutron-p and a proton-p \\ 
\qq 15  &   (0002\leftarrow 0002)  &      scattering of two neutron-p \\ 
\qq 16  &   (1100\leftarrow 1100)  &      scattering of a proton-ph pair \\ 
\qq 17  &   (1001\leftarrow 1001)  &      scattering of a proton-h-neutron-p pair \\ 
\qq 18  &   (0110\leftarrow 0110)  &      scattering of a neutron-h-proton-p pair \\  
\qq 19  &   (0011\leftarrow 0011)  &      scattering of a neutron-ph pair \\ 
\end{tabular}
}
\end{center}
\vspace{1.50cm}

Notice that this association is biunivocal because we are making the correspondence of
one independent sp-state with one value of each index, therefore the different
combinatorial sets of sp-states (nuclear configurations defined by a set of excitons
states) will be represented by different sets of indices and all possible configurations
will be decribed.

{Here the problem of ``large numbers" becomes apparent. For example, if we consider a
typical model space with \mas{ph}=\mas{nh}=28 available sp-states for holes and
\mas{pp}=\mas{np}=9 available sp-state for particles, the number of possible
configurations may be very large for large \na. One would have nearly 24
million total configurations
for \fe{56} excited at 60 MeV, and the number of possible scatterings of pairs of
excitons, to describe transitions in which \delc\na=0, would possibly involve up to
10\up{12} different combinations and the estimated time for direct computation, using
personal computer, would become of the order of 10\up{5} secs or more to obtain all
the TST listed in Tables
\iftth
\ret{_I}
\else
\ref{tablen.1}
\fi
and
\iftth
\ret{_II}
\else
\ref{tablen.2}
\fi
as a
function of \uc, \embm{for a given \na}.}

One simplifying aspect of the problem is that, according to the general expressions of
the \DMF\ the matrix elements of the residual potential (see \ree{3.1}) can be calculated
independently for each combination of initial and final nuclear configurations and do not
depend on the properties of the nuclear system as a whole, except for the self-consistent
nuclear mean field. This independence would be compromised if rotational and vibrational
nuclear modes were to be considered as part of the potential energy of sp-states.

The mean-field is assumed to be local and to include the effects of rotation and
vibration of the entire system phenomenologically, as explained below, with all
interactions completely represented by the adopted basis of sp-states.

For example, the term of the pre-equilibrium Hamiltonian that increases the number of
particles and holes by 2 is,\cite{fbgarXiv1,obp}
\bn{ \rbk{\aus+\a\aus+\b\bus+\d\bus+\g}{\bs\g\bs\d\as\b\as\a} =
 \sumss{12}{UM} \sums{\a\b\d\g} \absu{\vcs{\a\b\g\d}}{2}
 \dbokt{\pa\ha}{\aus+\a\aus+\b\bus+\d\bus+\g}{\pp\hp}{\bs\g\bs\d\as\b\as\a}
\; }
\bm{ = \sums{1}\sums{\sc\ta} \eu{\brak{\uc\mc}} \sums{\a\b\d\g} \absu{\vcs{\a\b\g\d}}{2}
\da\brac{\pa-2\dif\a\b,\ha-2\dif\d\g,\su,\ta} \;, }
where \absu{\vcs{\a\b\g\d}}{2} depends only on the sp-states being created or destroyed,
in this case sp-states (\a,\b,\d,\g) were not present in the initial configuration and
were added to the final one, but \vcs{\a\b\g\d} does not depend on \uc\ or (\pa,\ha), it
is not a direct function of the particular configurations \ket{\pa,\ha,\uc} and
\ket{\pap,\hap,\ucp} in which the
transition occurred, only indirectly through the sp-basis.
Therefore, for given exciton numbers and \uc\ the number of matrix elements that need to
be calculated decrease by a factor of
\da\brac{\pa-2\dif\a\b,\ha-2\dif\d\g,\uc,\mc}, with similar results for the other
microscopic transitions.

On the other hand, if for each type of microscopic transition \jas{am} is the maximum
value of the sp-states angular momentum and it is no greater than a given maximum \nas{0},
in units of \hbar/2 (for example \nas{0}=20),
and use the above definitions for \mas{\pa\ha}, \mas{\na\ha}, etc., then the total number
of elements of the \vcs{\a\b\g\d} matrix that need to be stored is \embm{smaller} than,
\bm{  \das{\max}=\nas{0}\dasu{b}{4} \;,
}
where \das{b}=\mas{\max}+1,
\mas{\max}=\max(\mas{\na\ha},\mas{\pa\ha},\mas{\na\pa},\mas{\pa\pa}),
\das{1} = \nas{0}\das{b},
\das{2} = \das{1}\das{b} = \nas{0}\dasu{b}{2},
\das{3} = \das{2}\das{b} = \nas{0}\dasu{b}{3} and
\das{\max} = \das{3}\das{b}. This is essentially equal to the
procedure used to represent integers using only the digits from 0 to 9, in terms of
powers of ``10", which in this case has been replaced by powers of ``\das{b}" with \nas{0}
replacing \das{b} in the ``zeroth power".
Therefore, for each sequence of 4 sp-states, corresponding to one microscopic transition,
with indices given by the parameters defined in \see{-5} to \see{-2} one can define an
index
\bm{      \nas{ind}=\nas{\ia2}+(\ias{2}-1)\times\das{\max} \;,}
where
\bn{
\nas{\ia2} = (\nas{1}\nas{0}+\nas{2}\das{1}+\nas{3}\das{2}+\nas{4}\das{3}+\jas{am}) }
\bm{
           = (\nas{1}\nas{0}+\nas{2}\nas{0}\das{b}+\nas{3}\nas{0}\dasu{b}{2}
+\nas{4}\nas{0}\dasu{b}{3}+\jas{am}) }
and \ias{2} indicates the different types of microscopic transitions, as in
\iftth \ret{Table_I}\else \ret{Table I}\fi
~and
\iftth \ret{Table_II}\else \ret{Table II}\fi
~above.

Because \brar{\nas{1},\nas{2},\nas{3},\nas{4}} are supposed to be no
greater than \mas{\max} it is clear that

\vst
\hstp\hstp
\max\brar{\nas{1}\nas{0}+\jas{am}}= \max\brar{\nas{1}}\nas{0}+ \max\brar{\jas{am}}=
\mas{\max}\nas{0}+ \nas{0} = \das{b}\nas{0} = \das{1}
\vst

\hst similarly
\max\brar{\nas{1}\nas{0}+\nas{2}\das{1}+\jas{am}}= \das{2},
\max\brar{\nas{1}\nas{0}+\nas{2}\das{1}+\nas{3}\das{2}+\jas{am}}=  \das{3} and

\vst
\hstp\hstp\hstp
\max\brar{\nas{1}\nas{0}+\nas{2}\das{1}+\nas{3}\das{2}+\nas{4}\das{3}+\jas{am}}=
\das{\max}.
\vst

Therefore, \nas{\ia2} in \see{-1} is never greater than \das{\max}, the various indices for
transitions with different ``\ia2" belong to different ranges of values of \nas{ind} and
the relation between these indices and the possible microscopic transitions is
biunivocal.

Then, the matrix elements can be stored in repository files, as a function of ``\ia2",
instead of being repeatedly calculated for each new TST.
The sequence of indices defining the possible configurations and their
enegies, for each \na, can also be calculated and stored in an series of independent
input files that will be read only when the \ket{\initial} configuration has \na\
excitons, where \ket{\initial} is the nuclear configuration before the microscopic
transition takes place.

Typically storage files containing all possible nuclear configurations for a given \na,
may have 10\up{7} lines or more depending on the maximum excitation energy, the number of
excitons and the spacing between levels of the sp-basis. In these cases the total number
of microscopic transitions is of the order of 10\up{14}, but using the above simplifying
rules and definitions of indices, these calculations can still be performed using
personal computers with not too long running times.

\subsubsecto{Nuclear Rotational and Vibrational modes}

If the potential energies for the vibration and rotation of the whole nuclear system have
not been taken into account in the definition of the sp-basis, the potential energy of
the total system, in its rotational and vibrational modes, can be exchanged with the
individual sp-states being created or destroyed during the microscopic transition and
originate a \ket{\final} with excitation energy \embm{different} from \ket{\initial}.
On the other hand, for a given maximum excitation \ucs{\max}, defined, for example,
by the incident energy in proton or neutron induced reactions, the energies of
\ket{\initial} and \ket{\final} configurations associated with each microscopic
transition must not be greater than \ucs{\max}.

The non conservation of the excitation energy at each microscopic transition, to account
for exchanges with dynamical modes of the whole system that are missing in the potential
energy of sp-states, can be
taken into account approximately by using a range of possible energies for each
microscopic transition, \ucs{\final}=\ucs{\initial}+\delc\uc, with \delc\uc\ defined
phenomenologically.

The necessity of a phenomenological \delc\uc\ is a another deficiency, or inconsistency,
of the present version of the \DFM, but it can be eliminated straightforwardly by
including potentials to describe the coupling of the rotational and vibrational modes of
the nuclear system to the movement of the sp-states,
in addition to the mean-field generated by the microscopic interactions among nucleons. Such
potentials were not included in the present work.

The eigenfunctions of the H.O. basis are given by known expressions in terms of
hypergeometic function.\cite{brussbook} The matrix elements \vcs{\a\b\g\d} are defined in
\trans\ by direct computation of the integrals of the wave-functions (Hermite and
Laguerre integrations over the configuration space) of the exciton states coming in and out
of the microscopic transition,
in the center of mass (CM) system, and using the Green function of the \embm{non
interacting} two-body Hamiltonian for the propagation of the pair of ``colliding"
excitons before and after the transition,\cite{mess61}
\bm{ \frac{ \eu{ik\abs{ \vec{\ras{1}}-\vec{\ras{2}} }} }{ \abs{\vec{\ras{1}}-\vec{\ras{2}}} } =
\ka \sumsu{\la=0}{\infty} (2\la+1) \jas{\la}(\ka\ras{<}) \hasu{\la}{(+)}(\ka\ras{>}) \pcs{\la}(\cos\a) \; }

\vhst in accordance with the IPM hypothesis.\cite{fbgarXiv1}

\subsecto{Numerical results}

The TST as a function of (\na,\uc) tend to increase for given \na\ and increasing \uc,
because the state density (degeneracy of nuclear configurations) tends to be larger for
larger \uc, although they are not strongly dependent on \uc.
This is a consequence of the mathematical definitions, for example,
Eq.\see{-4}.

In addition,  the maximum value of \uc, \ucs{\max}, in part defines the
``size" of the model space because, from \ree{4.2}, the number of ``hole" sp-states and
the maximum ``hole"  sp-energy may vary with \uc, depending on how realistic the
description of the sp-states is, although the number of ``particle" sp-states is always
fixed by the nuclear mass.
Then, depending on the description, if one considers arbitrarily large excitations the TST
could increase arbitrarily and become meaningless.

A simple phenomenological solution is to consider a
sp-energy cutoff, \eps{\cut}, to set a maximum for the hole energies, that would both
limit the size of the model space and yield physically meaningful TST  independently of
the specific description of the sp-states.

The necessity of a cut-off energy for sp-states is a third deficiency of the \DFM, as
this maximum should be deducible from the general basic assumptions of the model. This
is a less important deficiency though than the one regarding the
non-possibility of solution of the nuclear many-body problem and the \embm{necessity} of
an approximate basis, because one expects that for high enough energies the ``hole-state"
would in fact be in the continuum and the transition would be an ``emission",
not important for the definition of the TST.

Therefore, a parameter like \eps{\cut} is physically expected from a realistic description of
the nuclear mean-field and the inconsistency could be resolved by adopting a more
realistic nuclear potential, like the Woods-Saxon one, instead of the
H.O., with a well defined upper limit for sp-energies above which lies the continuum
spectrum.\cite{ring_shuck,wsax54}

The TST measure the strength of each microscopic interaction and
are directly connected with the observed cross sections.
They should have well defined values independent of \ucs{\max}, as the cross
sections for a given reaction cannot depend, for example, on the energy of the incident
particle, when all physical processes have been considered.

In this work the numerical results were obtained using the H.O. basis and an arbitrarily
fixed \eps{\cut}, independent of \ucs{\max}. The latter being used to limit only the
maximum energy of \ket{\initial}. The energy of \ket{\final} was also limited by
\delc\uc, as eplained in \ref{section.4.1.1}.

\subsubsecto{Comparison with \tng\ and \OXM.}

In a previous work\cite{spfnotes}, we presented the \tng\ estimates of the cross sections
of some p-induced reactions on \fe{56} and here we assess
how the \DFM\ affects the transition rates (TR) and cross sections for the same reactions.

The comparison of the TST with the transition rates (TR's) of \tng\ or \oxm\cite{kon04}
show important differences, in particular the TST corresponding to transitions that
increase \na\ by 2 have
relatively too large variations as a function of \uc, being too small for low \uc\ but
similar to the TR's at maximum.

Taking the results of \tng~ as reference,
this suggests that either the TST are not correctly defined in the \DFM, or the
numeric approximations used in TRANSNU are inadequate or the TST represent, in fact, an
improvement over the traditional definitions of \tng\ and the \OXM.

The TR's of \tng\ or \oxm\ have similar behavior, although not exactly the same
functional form as functions of (\na,\uc). They are defined by similar
phenomenological relations in both cases, but
in \tng\ the transition parameters are based
on the model developed by Kalbach \cite{k72,k8586} and correspond to a more elaborated
version of the ``standard" \oxm\
than the model presented in Ref.\cite{kon04}.

In general, the TR calculated by \tn\ or the \EXM\ show strong dependence
on (\na,\uc), but the functional form does not vary too much for increasing \uc, as we
 see in \ref{figure.1}, where \lams{+} and \lams{-}
are the rates for transitions that increase or decrease \na\ by 2,
respectively.

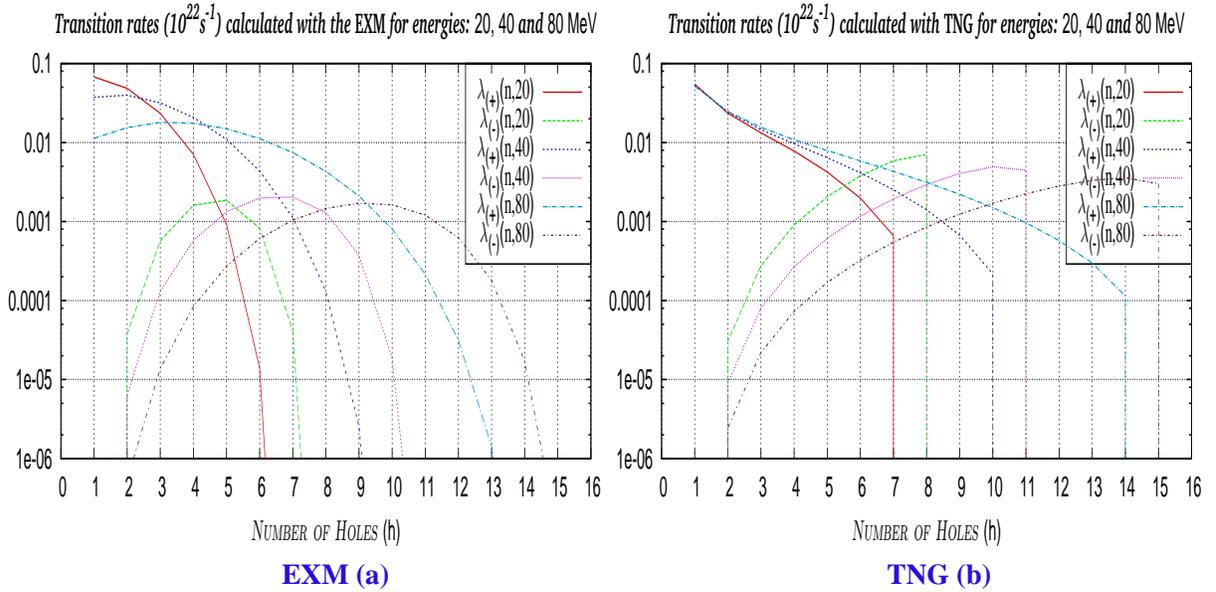
\makfigmm{1}{mixf-pics-oxma-tngb-ok-function-of-h-tg}{400}{Transition rates
calculated in the \EXM\ and by \tn\ as a function of the number of holes for excitation
energies 20, 40 and 80 MeV.  The functions are smooth, but show strong dependence on
(\na,\uc).
}{-1.1cm}{-2.5cm}{}

The \lams{+}(\na,\uc) of the \oxm\ in \ref{figure.1}(a) vary smoothly along
the complexity chain, tending to have a single well defined global maximum as a function
of \na, for high excitations,
while the \lams{+}(\na,\uc) of \tng, Fig.\ref{figuren.1}(b), are decreasing functions of
\na\ for all \uc.

In both cases the most probable exciton number is approximately given by \widetilde{\na}
=\brac{\ga\uc}\up{1/2}.\cite{gri66}

As we saw in \ree{3.5} and \ree{3.10}, \lams{+}(\na,\uc) is
proportional to the density of nuclear states after the transition and to the TR for the
reversed process,
\bm{ \frac{\lams{+}(\pa,\ha,\uc)}{\lams{-}(\pa+1,\ha+1,\uc)}  =
 \frac{\om(\pa+1,\ha+1,\uc)}{\om(\pa,\ha,\uc)} = \ras{\om}(\na,\uc)
\;. }
Therefore, the total transition rate for configurations belonging to a given exciton classs,
\lamc(\na,\uc) in \ree{3.4}, is a function of \ras{\om}(\na,\uc) and approximately a
function of
\lams{+}(\na,\uc)(1+1/\ras{\om}(\na,\uc)). The ratio \ras{\om}(\na,\uc) decreases rapidly
from one class to the next and becomes negligible for \na\ larger than \widetilde{\na}.
Consequently, the term
(1+1/\ras{\om}(\na,\uc))
becomes very large for large \na, favoring the time variations of occupation
probabilities with \na\ smaller than \widetilde{\na}.

This strong dependence of \lamc(\na,\uc) on \lams{+}(\na,\uc)(1+1/\ras{\om}(\na,\uc))
is very important for the \pe-emissions because, if \lams{+} is large, the occupation
probabilities, \pcs{\nu}(\na,\uc,\ta) in Eq.\ree{3.3}, tend to be larger for low \na\ at
the moment of emission.

In fact, in \tng, the \pcs{\nu}(\na,\uc,\ta) are assumed to be very high for low \na, i.
e. close to 0.5, for \na=1, for protons and neutrons (and nearly zero for alphas), with
the probabilities of classes that have \na\ close to \widetilde{\na} tending to increase
at each successive iteration of \ree{3.3}.
Due to the relation between \lams{+}(\na,\uc) and \lams{-}(\na+1,\uc) in Eq.\see{-0},
\tng\ assumes that a \pe\ emission happens when the ratio of \pcs{\nu}(\na,\uc,\ta)
between successive exciton classes becomes approximately equal to the corresponding
ratios of densities.  Then, if \lams{+}(\na,\uc) increases, the ratio of
\pcs{\nu}(\na,\uc,\ta) over \pcs{\nu}(\na+1,\uc,\ta) will increase and have
a greater chance to be equal to the (large) ratios of the corresponding nuclear
densities for low \na, and emissions in this region will be favored.

Having in sight that
\pe\ emissions at low \na\ tend to happen more frequently for lower energies, as
they correspond to the excitation of less particles from the ground state sp-levels,
larger \lams{+}(\na,\uc) tend also to favor emissions with lower \uc, therefore
increasing cross sections in this region of energies.

Due to the strong relation with \pe-emissions, the direct use of the
function \ras{\om}(\na,\uc) was introduced in the master equation \ree{3.3} of \tng,
giving additional consistency for the evaluation of \widetilde{\na} and
the emission rates, {\lamsu{\nu}{c}(\na,\uc,\eps{\nu}) in Eq.\ree{3.12}, as opposed to an
independent definition of these parameters.
This yielded smoother excitation functions in the regions where different exciton classes
contribute to the emission process.}

The corresponding parameters \lams{+} and \lams{0} calculated with TRANSNU are shown in
\ref{figure.2}.
Note that in \ref{figure.2}(a) \lams{0} is many orders of magnitude greater than \lams{+}
for all (\na,\uc) and \lams{-} is not calculated, in TRANSNU, in accordance with of the
``never-come-back" assumption.

The ratio of the parameters for ``intra-class", \lams{0}, and ``inter-class", \lams{+},
transitions,
\bm{ \ras{ex}(\na,\uc) = \frac{\lams{0}(\na,\uc)}{\lams{+}(\na,\uc)} \;,}
is usually \embm{greater than} ``1000", which is a ``threshold"
suggested by Pompeia and Carlson\cite{pc} to warrant the validity of
the exciton model description.
The large \ras{ex}(\na,\uc) obtained with TRANSNU would validate one of the basic
assumptions of the exciton model, that each exciton class can be considered as reaching
equilibrium, perfect configuration mixing, before emission or inter-class transitions,
but it is in conflict with the results of Ref.\cite{pc}.

The ratio \ras{ex}(\na,\uc) in \ref{figure.2}(a) does not vary importantly as a function
of \na, for a given \uc, which means that complete configuration mixing should occur at
all stages of the complexity chain, with no important difference for all energies.
This is in marked contrast with the results obtained with the H.O. basis, in
Fig.\ref{figuren.2}(b), where the \lams{0} become smaller than the \lams{+} for large
\na, due to a general relative increase of \lams{+} for low \na, especially for low
excitations. The discontinuity of the derivative of the TST, especially for low \uc, are
due to numeric precision and not related to definition of the functions.
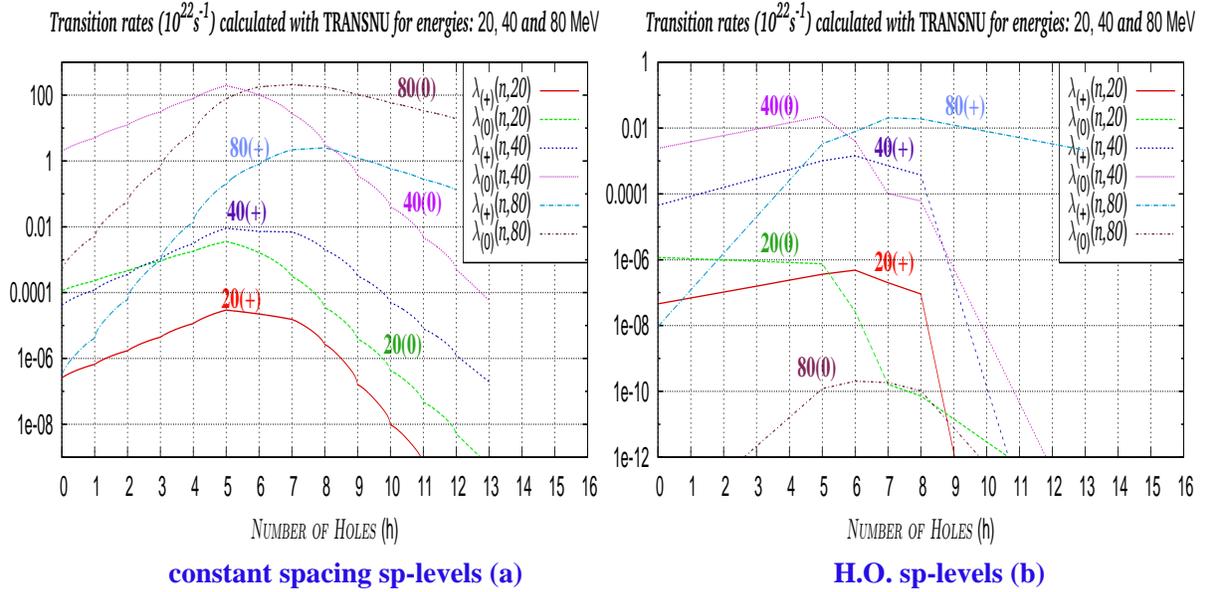
\makfigmm{2}{mixf-pics-transnu-lambda-plus-zero-const-spac+ho31-tg}{400}{Transition rates
\lams{+} and \lams{0} calculated by TRANSNU as a function of the number of holes
for excitation energies 20, 40 and 80 MeV. The functions are smooth and show strong
dependence on (\na,\uc), but \lams{+} has more pronounced variation with (\na,\uc) than
the corresponding \tn\ or \exm\ functions.  }{-1.1cm}{-2.5cm}{om-lambtng.jpg}

The \lams{+}(\na,\uc) of \tran\ in \ref{figure.2} show a more complicate behavior
as a function of \na,
than the TR's \tng\ or \OXM, resulting in part from the {influence of the structure of
excited nuclear levels} on the evolution of the complexity chain, but mainly the
microscopic interaction among nucleons.

This influence is even clearer in the dependence of the TST with \uc, as we see in
\ref{figure.3}, especially for the H.O. based estimate.
The decreasing magnitudes for high energies indicate only the decreasing
{number of combinations due to the limitations imposed on the model space.
It is not present in the TR's of the phenomenological approaches.}

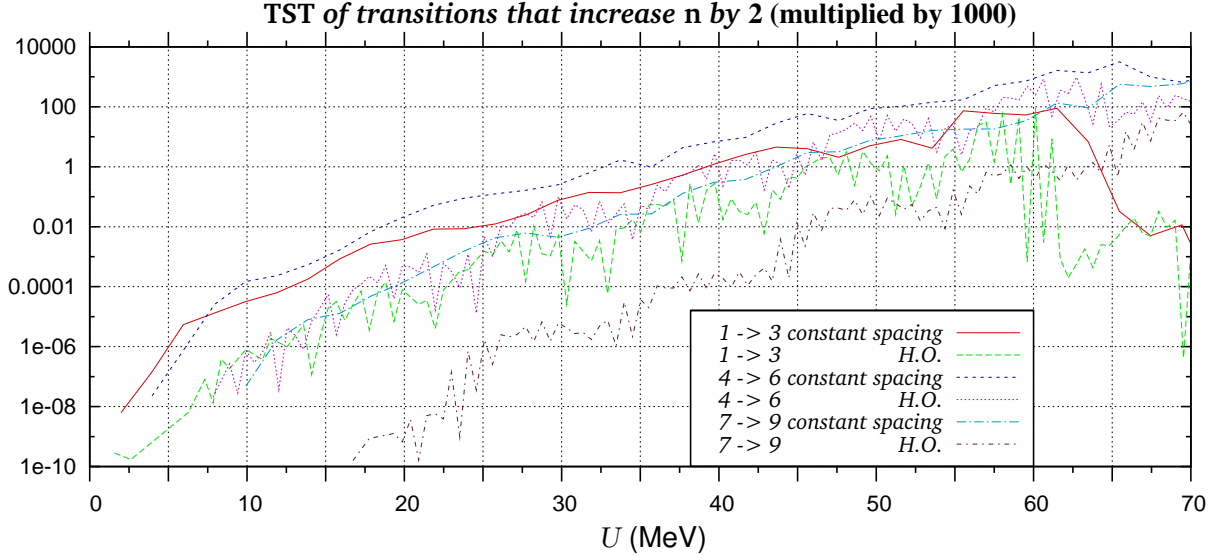
\makfigmm{3}{compare-transnu-tst-31-with-ho31}{400}{Transition strengths of
TRANSNU as a function of \uc, multiplied by 1000 to distinguish the curves more clearly,
for transitions that increase \na\ by 2, for \na=1, 4 and 7. The curves for \na=3, 5 and 6
are very close to \na=4, with similar ``plateaus" in the same regions.}{-1.1cm}{-2.5cm}{}

The mean-field parameter of Fermi's Golden Rule, term
\abs{\dbok{\na+2}{V(\uc)}{\na}}\up{2} in Eq.\ree{3.5}, can be defined as the sum of all
expected values of the operator of microscopic transitions, for given (\na,\uc), or the
mean potential energy for transitions without emission. It has
relatively low magnitudes
and non uniform variation, with the formation of approximate ``plateaus" around 20, 30,
40 and 50 MeV and increasing magnitude in between these regions. The ``plateaus" are more
well defined for larger \uc, as we see in \ref{figure.4} for the H.O. based model space.

On the other hand, as we see in \ref{figure.5},
the TR's of \tng\ or \OXM\ vary smoothly with \uc\ and
the relatively high magnitudes of the TR's for low \na, in comparison
with the corresponding TST, does not reflect the fact that in these regions the density
of excited states is expected to be low in nuclear systems and, therefore, the TR's
should increase with \na, at least in the region close to zero.

The presumed large density for low \na\ of the phenomenological models is an
unjustifiable heritage
of the old theory of metals, i. e. the approximate assumption of a ``highly degenerated
Fermi gas", independent of \na,\cite{fbgarXiv3}, and it should be discarded.
The TR's of \OXM\ only show the expected dependence with \na\ for high excitations, while
the \tng\ parameters do not have this dependence for any value of \uc.

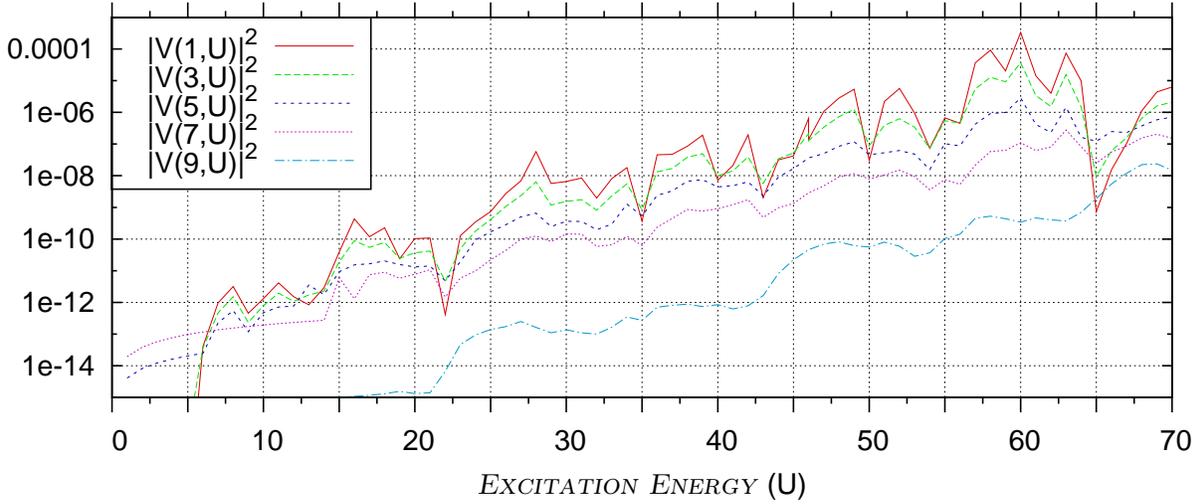
\makfigmm{4}{mfield-from-interpout12-ho31-case}{400}{Nuclear mean-field
for transitions that increase \na\ by 2, $|\!\!<\!n+2|V(U)|n\!>\!\!|$\up{2}, calculated
by TRANSNU, for the H.O. based model space, for various number of
excitons.}{-1.1cm}{-2.5cm}{}

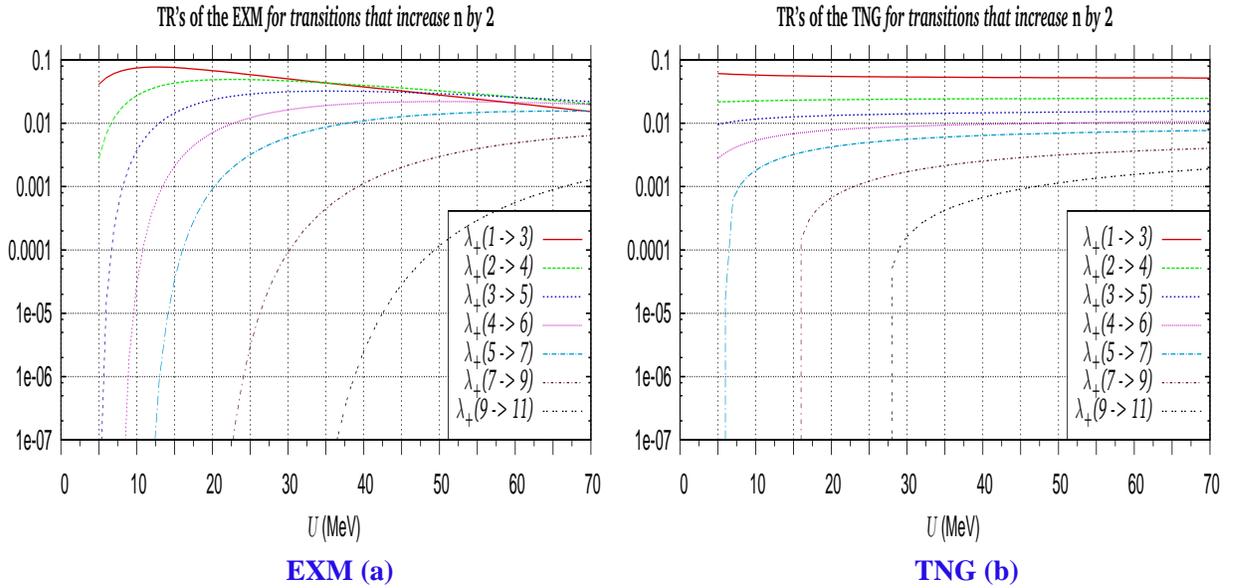
\makfigmm{5}{mixf-pics-compare-oxma-tngb-trs}{400}{Transition strengths of the \OXM\ and
\tng\ as functions of \uc, for transitions that increase \na\ by 2, for \na=1, 2, 3, 4
and 5.}{-1.1cm}{-2.5cm}{}

The low values of the TST for low \na\ reflect the fact that the density of
states grows quickly in this region, due to the fast increase in
the number of configurations for increasing number of excitons.
Therefore, the TST have a more realistic description of the low \na\ region than the
usual phenomenological approach.

The non cumulative densities of levels for the \OXM\ are shown in \ref{figure.6}. The
corresponding functions for \tng\ are very similar. 
They have no oscillations, because the statistical models assume no structure of
nuclear levels, and increase rapidly for low \uc, tending to flat for high energies.


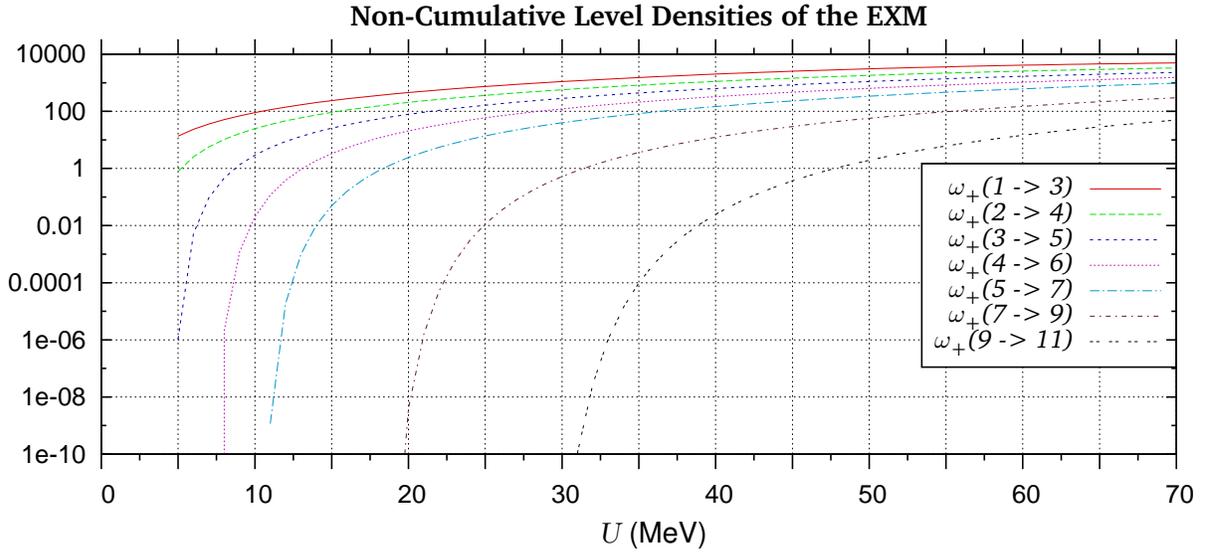
\makfigmm{6}{compare-oxm-levdnc-nts1-ttn-11p0e+03-timlofromhi}{400}{Non cumulative
nuclear level densities for transitions that increase \na\ by 2 in the \OXM,  as a
function of \uc, for \na=1 to \na=5.}{-1.1cm}{-2.5cm}{}


The corresponding functions calculated by TRANSNU have oscillatory behavior and tend to
decrease for high energies, due to \eps{\cut}, as we see in \ref{figure.7}.
In this case, the non-cumulative density of nuclear levels results directly from 
combinatorial calculations.

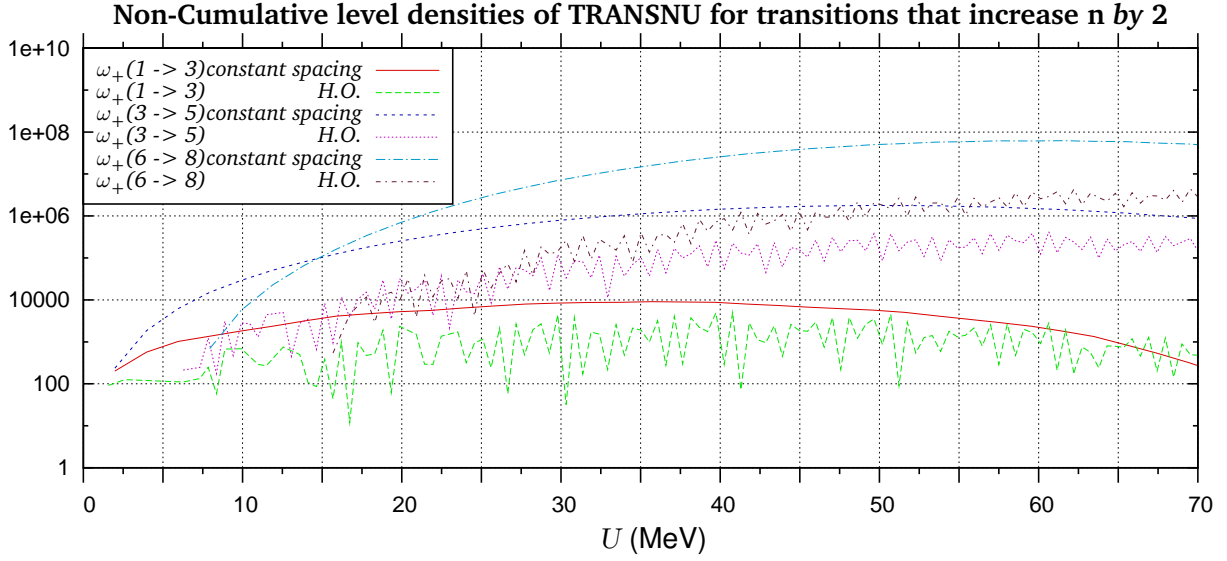
\makfigmm{7}{compare-transnu-levdnc}{400}{Non cumulative nuclear level densities
calculated by TRANSNU as a function of \uc, for transitions that increase \na\ by 2, for
\na=1, 3 and 6.}{-1.1cm}{-2.5cm}{}

\subsubsecto{Comparison with EXFOR}

The cross sections for \pa-induced reactions on \fe{56} are shown in the following
figures.
The main procedure of \tng\ was used for the evaluation of the cross sections and the
functions for the TR's of the \OXM\ and the TST of TRANSNU were calculated independently
and used as input of \tng.

The cross sections calculated with the \tn\ model and the independent definition of the
\OXM\ of Ref.\cite{kon04} are shown in Figs. \ref{figuren.7} and \ref{figuren.8},
respectively.

The the two estimates are approximately the same despite the differences of the
corresponding \lams{+}(\na,\uc), as we saw in Figs. \ref{figuren.1} and  \ref{figuren.5}.
These differences become important only for a detailed description of the excitation
functions, for smaller ranges of \uc.

After the consistent redefinition of \lams{-} is made, from \lams{+}, using
\ras{\om}(\na,\uc) in the master equation, the resulting cross sections 
obtained with \tng\cite{tng,k8586} or the \OXM\ of \cite{kon04} are almost identical,
indicating that the general behavior of the TR's and their magnitudes, whether
oscillatory or uniform as functions of (\na,\uc), are more important than the specific
functional dependence.

The same cross sections calculated with the \lams{+}(\na,\uc) of \trans\ are given in
Fig.\ref{figuren.9}, for the constant inter-spcing basis and Fig.\ref{figuren.10} for the
H.O.  basis. Although the estimates are not as good as the phenomenological ones they
have a reasonably correct description of the experimental cross sections
for the activation energy, the local maxima and average magnitude.

It is mainly the magnitude for small regions of \uc\ that is not well defined by the present
version of the \DFM, due to the different dependence of the TST with the energy, for
the various \na\ in comparison with the \EXM\ or \tng\ functions, as we see in
\ref{figure.3} and \ref{figure.5}, respectively.

The TST tend to oscillate with energy, while the phenomenological TR's have uniform
dependence. The uniformity means that the ratio of TR's for reversed processes,
\ras{\om}(\na,\uc) in Eq.\ree{4.12}, is approximately the same for energies close to each
other and \delc\na=0,\pm2. Therefore,
\widetilde{\na} will also be a uniform function of \uc\ and the contributions for
emission of the various exciton classes will be well defined, for each small region
around a given \uc, with excitation functions determined by the activation energies of
the various residual nuclides and the magnitude of \lams{+}, as explained in the analysis
of
\ras{\om}(\na,\uc).

In the case of an oscillating 
\widetilde{\na}(\na,\uc) the contributions of neighboring energies will not be uniform
and the correct description of the excitation functions will be more directly dependent
on the magnitude of \lams{+}(\uc), to compensate for the non uniformity of
\widetilde{\na}(\uc), espcially for low \na.

\newpage
\makfigmm{8}{mixf-fe56-nts0-default-factim1-09-16-13}{400}{
Cross sections for \pa-induced reactions on \fe{56} calculated with
\tn.}{-1.7cm}{-2.5cm}{}

\newpage
\makfigmm{9}{mixf-fe56-nts1-ttn-11p0e+03-timlofromhi}{400}{
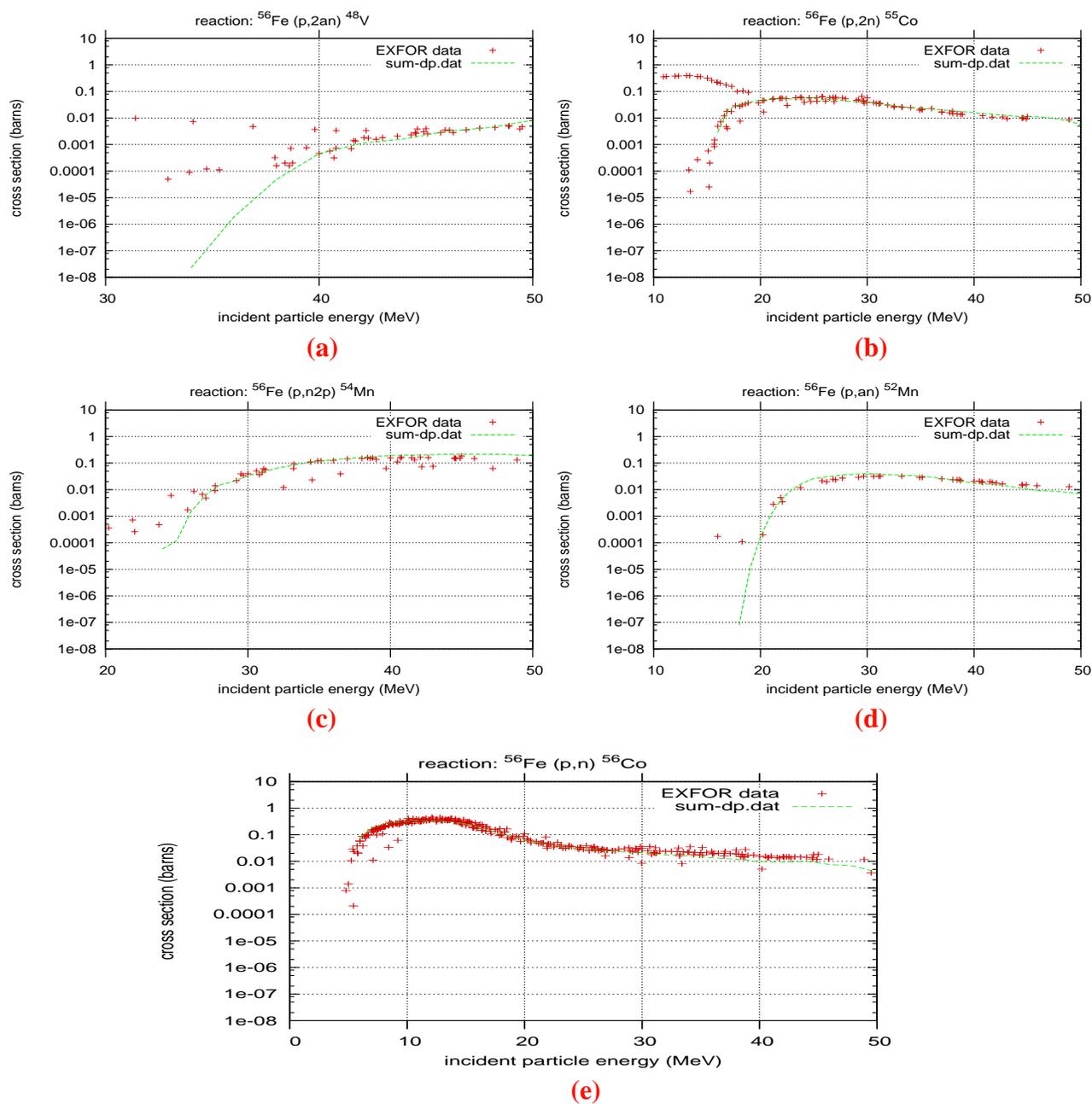
Cross sections for \pa-induced reactions on \fe{56} calculated with the \OXM\ formulation
of Ref.[14].}{-1.7cm}{-2.5cm}{}

\newpage
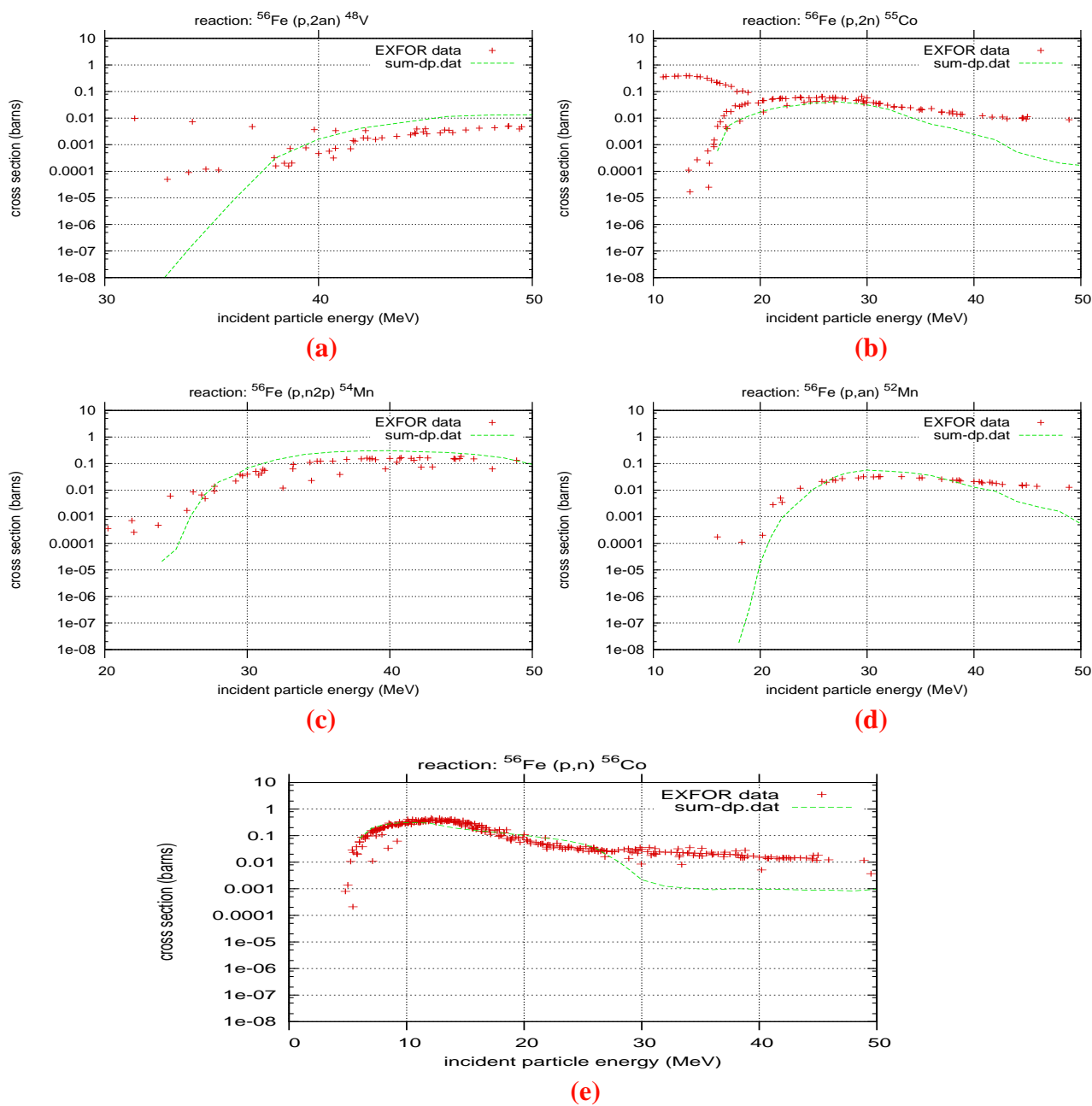
\makfigmm{10}{mixf-fe56-nts2-ttn-4p0e+06-newnx}{400}{ Cross sections for \pa-induced
reactions on \fe{56} calculated with the TST obtained with the constant interspacing
between sp-levels in the model sapce of \tran.
}{-1.7cm}{-2.5cm}{}

\newpage
\makfigmm{11}{mixf-fe56-nts2-ttn-4p0e+06-ho31-newnx}{400}{ Cross sections for \pa-induced
reactions on \fe{56} calculated with the TST obtained with the H.O.
sp-levels in the model sapce of \tran.}{-1.7cm}{-2.5cm}{}

\newpage
\secto{Final Comments and Conclusion}

We made a simplified presentation of some important aspects of the \DFM\ and the \tng\
model code to show their most relevant features and differences,
to compare the results of model codes \trans\ and \tn\ for the evaluation of \pe-emission
cross sections.

The cross sections calculated with \tn\ have been shown to be equivalent to those
obtained with an independent definition of the \OXM\ given in Ref.\cite{kon04}, as we see
in Figs. \ref{figuren.8} and  \ref{figuren.9}.

We used the strong dependence of the \pe\ emission on the parameter
\ras{\om}(\na,\uc) of Eq.\ree{4.12}, in \tng, to redefine \lams{-} consistently from the
\lams{+} in the master equation \ree{3.3} and obtain smoother excitation functions in the
regions where different exciton classes contribute.
The resulting p-induced cross sections on \fe{56} obtained with \tng\cite{tng,k8586} or
the \OXM\ of \cite{kon04} are almost identical, indicating that the general uniform
behavior of the TR's and their magnitudes as a function of (\na,\uc) is more important
than the specific functional dependence, as exemplified by the functions plotted in
\ref{figure.1}. 
The specific dependence is important only for the {detailed} description of the
excitation functions for each small region of \uc, because it affects \ras{\om}(\na,\uc).

On the other hand, the parameters of TRANSNU, the TST, tend to vary more pronouncedly
and non uniformly than the corresponding phenomenological ones,
especially as a function of \uc\ and for low \na, being very low for low \uc. This is the
main cause of the large differences with experimental cross sections obtained with the
parameters of TRANSNU, very close to regions where the excitation function is reasonably
well estimated.

We saw that due to the strong relation of \ras{\om}(\na,\uc) with \pe-emissions, the
oscillations of the estimated excitation functions are related to the incorrect
compensation of the oscillations of \widetilde{\na} and the TST as function of (\na,\uc).

The oscillating behavior of the
\lams{+}(\na,\uc) of \tran\ in Figs. \ref{figuren.2} and \ref{figuren.3} was determined
in part by similar oscillations of combinatorial origin of the nuclear level density of
favorable states, \oms{+}(\na,\uc), but mainly by the non uniformity of the total
expected value of the operator of microscopic transitions,
\abs{\dbok{\na+2}{V(\uc)}{\na}}\up{2} in Eq.\ree{3.3} and \ref{figure.4}, which is in
part due to lack of enough numerical precision of the present calculations.

The total mean-field associated with \lams{+}(\na,\uc) of TRANSNU has regions of nearly
flat dependence on \uc, which is completely absent in the phenomenological models. In
addition, it shows a steady increase for low \uc\ and low \na, which is in
agreement with the fact that nuclear densities tend to increase rapidly for low \na\ and
is physically more correct than the relativey large \lams{+} for low \na\ of the
phenomenological estimates. Therefore, in this respect, the TST represent an improvement
over the phenomenological TR's.



\subsecto{Deficiencies of the present version of the \DFM}


The definition of the model space in TRANSNU, either using the complete H.O. basis with
strong spin-orbit ``modulation" or the same single particle basis
except for the constraint that the sp-levels have fixed interpacing, are {essentially
phenomenological} as they are not obtained from the direct solution of the nuclear
many-body problem.
Therefore, it represents an intrinsic \embm{deficiency} of the present form of the DMF,
because it is not obtained directly from the microscopic definitions, as the formalism
prescribes.
This inconsistency with the original proposal of the \DFM\ cannot be avoided, but it can be
minored by using an approximate self-consistent definition of the nuclear mean-field, as the
Hartree-Fock approach.\cite{fetter}
Having in sight that strong nuclear forces are still not fully understood,\cite{jenk08}
the research of self-consistent approaches for the nuclear mean-field could be an
important and fruitful branch in the future developments of the \DFM.

The use of the H.O. basis, either with constant inter-spacing or strong spin-orbit
coupling, makes
the number of hole sp-states above \eps{\fc} in the nuclear ground-state \embm{infinite},
because the H.O. potential is an infinite parabolic well.
This particular inconsistency was avoided in this work by using an arbitrary cut-off
energy, \eps{\cut}, to define a maximum value for the hole sp-levels, \eps{\ha}, but it
could be solved in a more mathematically consistent way by adopting the basis of a finite
realistic potential, like the Woods-Saxon one.\cite{wsax54}

Even if the maximum hole energy is finite, the total nuclear excitation energy can still be
infinite, if the H.O. basis is used and the number of excitons is large enough, then
another energy cut-off, for the entire nuclear system, must be defined, \ucs{\max}, to
avoid infinite nuclear excitation. This is also a minor inconsistency of our present
calculations that could be avoided with a realistic nuclear mean-field potential.

Although \ucs{\max} naturally limits \eps{cut}, the two cutoffs are in fact independent
because the first determines the size of the model space and the number of hole states
contained in the range defined by the second.
The independent definition \ucs{\max} and \eps{cut} yields continuous TST as a function
of \eps{cut}, which is the physically expected behavior, while if only \ucs{\max} is
defined the TST, for given (\na,\uc), become a function of the cutoff.
The calculated TST are not strongly dependent on \eps{\cut} but tend to have increasing
magnitudes for increasing \eps{\cut} because the degeneracy of nuclear configurations
also increases when there are more hole sp-states available for transition.

Therefore, we used an independent definition \ucs{\max} and \eps{cut} in this work.

At last, a third inconsistency of the present version of the \DFM\ is related with the
dynamical modes of the system as a whole.  For a given maximum excitation \ucs{\max} the
energies of the configurations associated with each microscopic transition,
\ket{\initial} and \ket{\final}, must not be greater than \ucs{\max}, but the potential
energy of the total nuclear system, in its rotational and vibrational modes, can be
exchanged with the individual sp-states being created or destroyed during the transition
and possibly originate a \ket{\final} with excitation energy \embm{different} from
\ket{\initial}.\cite{okrice29}

The non conservation of the excitation energy, in the version of the \DFM\ used in this
work, was taken into account approximately by using a range of possible values for the
variation of energy for each microscopic transition,
\ucs{\final}=\ucs{\initial}+\delc\uc, with \delc\uc\ defined phenomenologically.
The necessity to consider a phenmenological \delc\uc\ can be solved by considering the
coupling between the sp-states and the collective modes of the total nuclear system in
the definition of the basis of the model space.

Therefore, only one inconsistency of the present version of the \DFM\ is unavoidable, and
the others should introduce corrections in the magnitudes and oscillatory behavior of the
TST as functions of (\na,\uc), in the various ``small regions" of \uc, defined
by the different sets of \na\ that importantly contribute to the excitation functions in
each region.

\subsecto{Conclusion}

Despite the inconsistencies of the version of the \DFM\ used in this work and the not
completely solved numerical problems, related mainly with precision of the calculations,
to eliminate the physically non meaningful noises in the strong oscillations of the TST,
especially with \uc\ and for low \na,
we consider the results presented in this work very promising.

We believe that the cross sections for the \pa-induced on \fe{56} are a good
estimate of the quality of the calculations that the present version of TRANSNU can
offer.

The possibility of a consistent redefinition of \lams{-} from \lams{+}, in the
determination of \widetilde{\na}(\na,\uc) in \tng, made the results of TRANSNU, for the
TST of \fe{56}(\pa,\xa) reactions, qualitatively correct.
In particular, we obtained good estimates for the average magnitudes of the cross
sections, for the majority of small regions of the excitation functions where the
phenomenological description is also well defined.

\newpage  
\addtocounter{sectionp}{1}
\setcounter{section}{\value{sectionp}}

\end{document}